\documentclass{aa}       % New LaTeX A&A Standard Fonts

\usepackage{txfonts}
\usepackage{graphicx}
\usepackage{amssymb}
\usepackage{longtable}
\usepackage{natbib}

\begin{document}

\title{Stability of three-dimensional relativistic jets: implications for jet collimation}

\author{M. Perucho\inst{1,2}, J.M. Mart\'{\i}\inst{2}, J.M. Cela\inst{3}, M. Hanasz\inst{4}, R. de la Cruz\inst{3}, F. Rubio\inst{3}}%\and
        
    %}

\authorrunning{Perucho et al.}

\titlerunning{Stability of 3D relativistic jets}

\institute{Max-Planck-Institut f\"ur Radioastronomie, Auf dem H\"ugel 69, 53121 Bonn, Germany 
\and
Departament d'Astronomia i Astrof\'{\i}sica. Universitat de Val\`encia. C/ Dr. Moliner 50, 46100 Burjassot (Val\`encia), Spain
\and Barcelona Supercomputing Centre, C/ Jordi Girona, 29, 08034 Barcelona, Spain
\and Toru\'n Centre for Astronomy, Nicholas Copernicus University, ul. Gagarina 11, PL-87-100 Toru{\'n}, Poland
}

\offprints{M. Perucho, \email{manel.perucho@uv.es}}

\date{Received date / Accepted date}

\abstract{
%%context
{The stable propagation of jets in FRII sources is remarkable if one takes into account that large-scale jets are subjected to potentially highly disruptive three-dimensional (3D) Kelvin-Helmholtz instabilities.}
%%aims
{Numerical simulations can address this problem and help clarify the causes of this remarkable stability. Following previous studies of the stability of relativistic flows in two dimensions (2D), it is our aim to test and extend the conclusions of such works to three dimensions.}
%%Methods
{We present numerical simulations for the study of the stability properties of 3D, sheared, relativistic flows. This work uses a fully parallelized code (\textit{Ratpenat}) that solves equations of relativistic hydrodynamics in 3D.}
%%Results
{The results of the present simulations confirm those in 2D. We conclude that the growth of resonant modes in sheared relativistic flows could be important in explaining the long-term collimation of extragalactic jets.} {}
\keywords{galaxies: jets – hydrodynamics – instabilities}}

\maketitle

\section{Introduction} \label{intro}

Extragalactic jets from AGN present a morphological dichotomy between FRI and FRII type jets \citep{fr74}. 
The former \citep[e.g., 3C~31,][]{lb02} show a disrupted structure at kiloparsec scales, whereas the latter 
\citep[e.g., Cyg A,][]{cb96} are highly collimated. Although the origin of this dichotomy is a complex combination 
of several intrinsic (e.g., jet power) and external (i.e., environmental) factors, the stable propagation of jets 
in FRII sources is remarkable considering that large-scale jets are subjected to potentially highly disruptive 
three-dimensional (3D) modes of the Kelvin-Helmholtz (KH) instability. Any perturbation may couple to an instability and 
grow in amplitude, becoming potentially disruptive \citep{pe+05}, hence the importance of studying their development and growth 
from the linear to the nonlinear regime in terms of the flow parameters.  

Jet stability in the relativistic regime has been thoroughly studied in different scenarios and parameter ranges 
\citep[see][for a recent review on the state-of-art]{ha06}, from both the analytical and numerical perspectives. The 
linear analysis of KH instability in relativistic flows started with the work of \cite{ts76} and \cite{bp76}, who derived 
and solved a dispersion relation for a single plane boundary between a relativistic flow and the ambient medium. Next, 
\cite{ftz78} and \cite{ha79} examined properties of the KH instability in relativistic cylindrical jets. The effects of a 
shear layer have been examined by \cite{ur02}. More recently, \cite{ha08} has studied the stability of magnetized relativistic 
jets with poloidal magnetic fields.

An extension of the linear stability analysis in the relativistic case to the weakly nonlinear regime has been performed by 
\cite{hz95,hz97} and led to the conclusion that KH instability saturates at finite amplitudes because of various nonlinear effects. 
The most significant effect results from the relativistic character of the jet flow, namely from the velocity 
perturbation not exceeding the speed of light in the reference frame of the jet. This result was confirmed by numerical 
simulations \citep{pe+04a,pe+04b}. 

In the purely nonlinear regime, \cite{ma97}, \cite{ha98} and \cite{ro99} demonstrate that jets with high Lorentz factors 
and high internal energy are influenced very weakly by the Kelvin-Helmholtz instability, contrary to the cases with lower 
Lorentz factors and lower internal energies. The former do not develop modes of KH instability predicted by the linear 
theory, which is interpreted as the result of a lack of appropriate perturbations (triggered by the backflow in this 
kind of simulations) generating the instability in the system, because in the limit of high internal energies of the 
jet matter the Kelvin-Helmholtz instability is expected to develop with the highest growth rate. 

A combination of linear and nonlinear studies of jet stability in the relativistic case was applied for the first 
time by \cite{ha98} in the case of axisymmetric, cylindrical jets and then extended to the 3D case by \cite{ha01} 
in the spatial approach and by \cite{pe+04a,pe+04b} in the temporal approach, for 2D slab and 
cilyndrical jets, including the case of sheared flows in \cite{pe+05}. Recently, \cite{mi07} have extended this 
study to 3D relativistic magnetized jets, which may be stabilized (with respect to current-driven modes) 
by a sheath layer surrounding the jet core. 

Focusing on the combination of linear and nonlinear studies, \cite{pe+04a,pe+04b} studied the growth of single symmetric 
modes in relativistic jets, sweeping different parameters, such as jet temperature, Lorentz factor, and density ratio, with 
the external medium. The linear work confirmed that KH modes grow faster in slower and hotter flows than in cold and fast 
ones. In Perucho, Mart\'{\i} \& Hanasz (2005, Paper I from now on), this work was extended to sheared jets to study the growth of 
a mixture of symmetric and antisymmetric modes in this case. In \cite{pe+04b} and Paper~I, the authors determined the stability 
properties of relativistic flows for a wide range of parameters in density contrast ($10^{-4}-0.1$), Lorentz factor ($5-20$) 
and specific internal energy ($0.08-76.5\,\rm{c^2}$). It was shown that the inertia of the jet, in terms of the density 
contrast with the external medium, the relativistic Mach number and the Lorentz factor are the  parameters that determine 
the stability of jets in the nonlinear regime, as a relativistic counterpart to the work of \cite{bm94} in the classical 
case. Moreover, in Paper I and \cite{pe+07} (Paper II, from now on), it was shown that the growth of high-order 
Kelvin-Helmholtz modes developing in the shearing layer, hereafter referred to as resonant modes, could dramatically change  
the nonlinear evolution of the flow in the case of cold and fast jets. 

The results could be summarized as follows. a) Cold and slow jets (small relativistic Mach number and Lorentz factor) 
are unstable with the growth of the instability and are disrupted after the generation of a shock front crossing the boundary 
between the jet and the ambient medium, which results from the development of long wavelength KH instability modes 
(UST1 jets). b) Hot and slow jets (intermediate values of relativistic Mach numbers and Lorentz factor) are also unstable, 
but disrupted and mixed in a continuous way by the growth of the mixing layer down to the jet axis (UST2 jets). c) Faster 
(high values of Mach number and/or Lorentz factor) jets develop short-wavelength, high-order modes that grow in the shear 
layer and saturate the growth of the instability without loss of either collimation or mixing, and generate a hot shear layer 
around the core of the jet (ST jets). d) Lighter jets are more unstable than denser ones (UST1). 

In this work, we extend the previous studies to 3D and show the existence of a new mechanism to prevent jet decollimation 
based on the development of resonant KH perturbations. With the aforementioned experience in mind, and taking into account 
that 3D simulations are highly demanding in terms of computational resources, we present here three numerical simulations 
using parameters that are representative of the different stability regions found in Paper I, but now including helical and 
elliptic modes. It is our aim to test whether the same conclusions are valid and, thus, if resonant modes provide a general 
stabilising mechanism in the absence of strong magnetic fields. 

The paper is organized as follows. Section~\ref{code} is devoted to presenting of a new 3D-RHD code for these 
simulations and the numerical setup used. Section~\ref{res} contains the results of the linear analysis for the parameters 
considered and the simulations. In Section~\ref{disc}, the results are discussed and the conclusions of this work are 
presented in Section~\ref{conc}. 

\section{Numerical simulations}\label{code}

This work was done with \textit{Ratpenat}, a new finite-difference code based on a high-resolution 
shock-capturing scheme that solves the equations of relativistic hydrodynamics in 3D written in 
conservation form. It is based on the 2D code used in all the series of works on jet stability performed in our 
group (see Paper I and references therein), and shares algorithm and structure with \textit{Genesis} \citep{alo99}. 
\textit{Ratpenat} has been parallelized using a hybrid scheme with both parallel processes (Message Passing Interface, MPI) 
and parallel threads (OpenMP) inside each process. This hybrid scheme is well adapted to the typical supercomputer 
architecture, i.e., a cluster of multi-processors. \textit{Ratpenat} can exploit all the memory available in a 
computational node and keep all the cores in the node busy with OpenMP threads. 

The parallelism between computational nodes is exploited with the MPI layer. Moreover, this hybrid scheme reduces the total 
amount of communications, because the number of MPI processes is reduced to the number of nodes and prepares the code for 
future supercomputer environments where load imbalance between MPI processes can be solved dynamically with the proper 
management of threads priorities. The distribution of computing load among the different participating processors is done 
by dividing the numerical grid in domains along the jet axis and distributing each portion to one of those processors. 
Communication is made before every temporal subcycle and the data provided are used as boundary conditions for each processor. 
The super-computer Mare Nostrum, at the Barcelona Supercomputing Centre (BSC), is composed by nodes that include four processors 
with shared memory. The code was tested to scale up to 256 nodes in Mare Nostrum, i.e., 1024 processors, with an appropriate 
grid size ($x \times y \times z = 256 \times 8192 \times 256$). 

The initial conditions in the simulations are similar to those used in Paper I for the 2D case: a portion of an inifinite jet, 
represented by periodic boundary conditions in the axial direction and outflow boundary conditions in the radial directions. 
The size of the grid is $16\,R_j$ (jet radii) along the axis and $8\,R_j$ transversally ($4\,R_j$ on each side of the jet axis), 
plus an extended grid up to $200\,R_j$ on each side, resulting in a grid with $512\,\times\,512\,\times\,512$ cells. The 
resolution is 32 cells$/R_j$ in the homogeneous grid. This is much less than that in the 2D simulations. The influence of 
the resolution on the results will be discussed in the next section.  
 
Each jet is initially in pressure equilibrium with the ambient medium. A shear layer is introduced between the jet and 
the ambient medium for axial velocity and rest-mass density:
\begin{equation}
v_z(r) = \frac{v_j}{\cosh(r^m)} \label{shearlayer1},
\end{equation}
\begin{equation}
\rho_0(r) = \rho_{0a} - \frac{\rho_{0a} - \rho_{0j}}{\cosh(r^m)}, 
\label{shearlayer2}
\end{equation}
where $r$ is the radial coordinate, $\rho_{0j}$ and $\rho_{0a}$ are the density of the jet and the ambient medium, 
respectively, and $m$ is the steepness parameter of the shear layer. In the 3D simulations, we used $m=16$, 
which implies a wider ($\sim 0.3\,R_j$) layer than in the 2D jets in Paper I ($\sim 0.2\,R_j$). Both the jet and the 
ambient medium are described as relativistic ideal gases with adiabatic index $\Gamma = 4/3$. Thus, we implicitly assume 
that the jet is surrounded by a hot cocoon. 

We chose three representative models out of those studied in Papers I and II: Model B05, D10, and B20, belonging to 
each of the three families mentioned in the Introduction. Their parameters are summarized in Table~\ref{tab:1} 
(see also \cite{pe+04a} and Paper I). The columns give, for each model, the Lorentz factor, the 
relativistic Mach number, the density contrast, the adiabatic index, the specific energy density of the jet and 
the ambient medium and the sound speed of the jet. Models B05 and B20 are shown along the paper as extreme cases, 
belonging to groups UST~1 and ST in the stability plane derived in Paper I (Fig.~21 in Paper I), so these models are 
taken as paradigms of those two groups. Model D10 shares properties of both B05 and B20, belonging to the 
UST2 group in the stability plane. Following the similarities for all models within these groups, we expect that any 
conclusion derived for the models presented here can be generalized to all the models falling in the same regions, 
including lighter, slower, faster, hotter, or colder jets depending on the group.

\begin{table*}{
\label{tab:1}
\begin{center} 
  \begin{tabular}{lcccccccc}
    \hline
    model & $\gamma_j$ & $M_j$ & $\rho_j/\rho_a$ & $\Gamma$& $\varepsilon_j (c^2)$ &  $\varepsilon_a (c^2)$ & $c_{sj} (c)$ &  resolution (cells/$R_j$)\\
    \hline
    B05 (3D)& 5.0 &  13.2 & 0.1 & 4/3 & 0.42 & 0.042 & 0.35 &  32 / 64$^a$ \\
    B05 (2D)& - &  - & - & - & - & - & - &  32 / 128x256$^b$\\
    D10 (3D)& 10.0 & 14.2 & 0.1 & 4/3 & 60.0 & 6.0 & 0.57 & 32 \\
    B20 (3D)& 20.0 & 54.0 & 0.1 & 4/3 & 0.42 & 0.042 & 0.35 & 32 \\
    B20 (2D)& - & - & - & - & - & - & - & 32 / 128x256$^b$\\
  \end{tabular}
\end{center}
\caption{Parameters of the simulations discussed in this paper. $^a$ This simulation was repeated (in the linear regime only) 
doubling the resolution. $^b$ The numbers correspond to the axial times radial resolution when it is not the same 
in all the dimensions.}
}
\end{table*}

A combination of linear pinching, helical, and elliptic sinusoidal perturbations is excited in all three simulations. 
These represent the fundamental mode of the numerical box ($\lambda=16\,R_j$) and the first three of its harmonics. 
The amplitude of the pressure perturbation is set by taking into account that the linear regimes are longer for colder and 
faster jets and that the initial amplitude does not change the long-term evolution of the jet (see Paper I).
The amplitude of the perturbation is thus greater in model B20 in order to avoid a very long linear regime 
\citep[see, e.g.][]{pe+05}. In Paper I it was proven that these perturbations couple to the fastest growing modes at 
the given wavelength and that any other shorter harmonics, such as the short, resonant modes, can be excited. Therefore, 
we have a representative sample of the different types of modes that grow in such a scenario and are able to test the 
general response of each set of parameters to the growth of the instability. 

The simulations were run for between 18 and 30 days in 32 nodes (128 processors) in Mare Nostrum (BSC), depending on the run. 
The number of nodes used was chosen on the basis of the grid dimensions, as explained above.

\section{Stability of three-dimensional relativistic jets}\label{res}

\subsection{Linear regime}
\begin{figure*}[!ht]
    \centering
\includegraphics[clip,angle=0,width=0.3\textwidth]{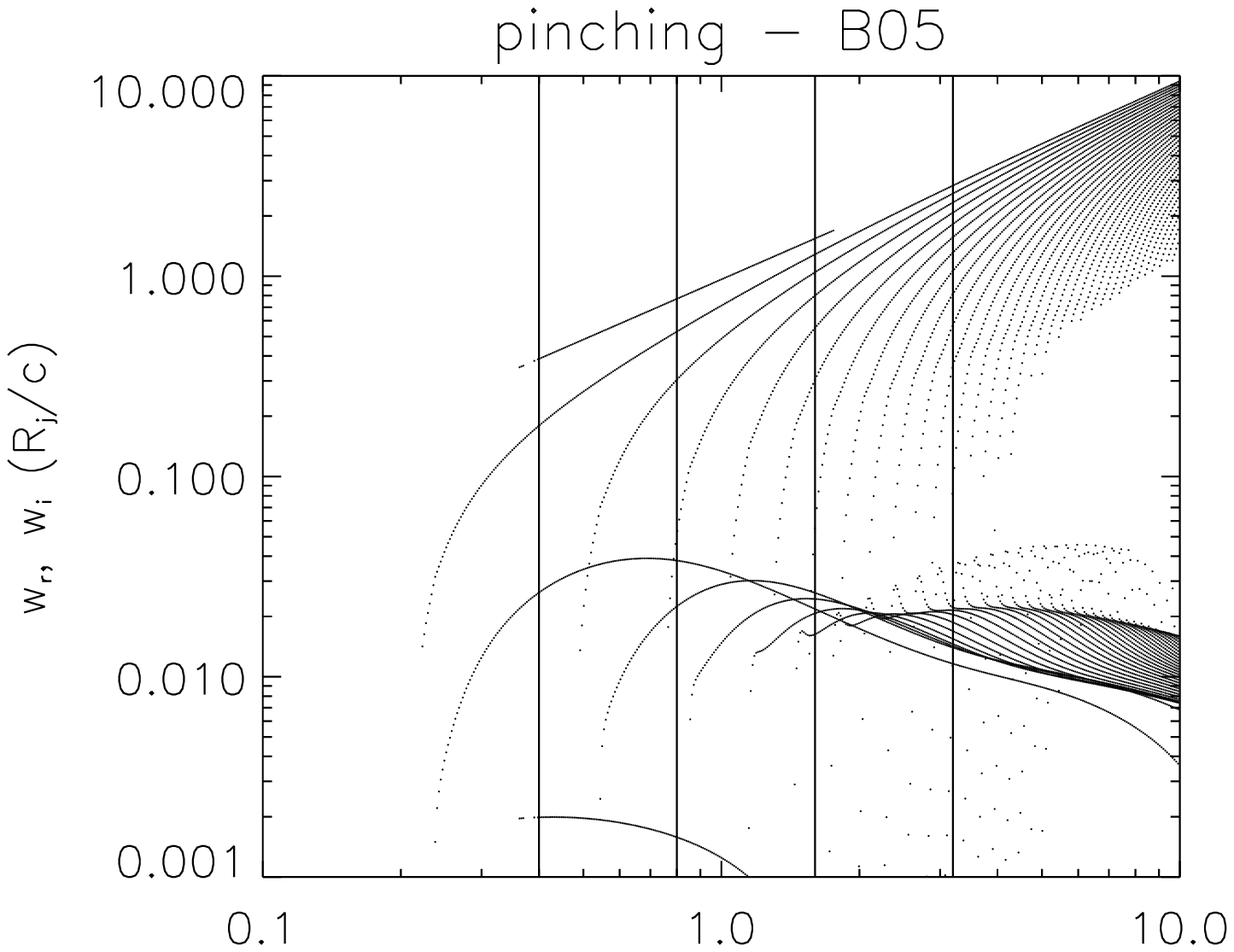}\quad
\includegraphics[clip,angle=0,width=0.3\textwidth]{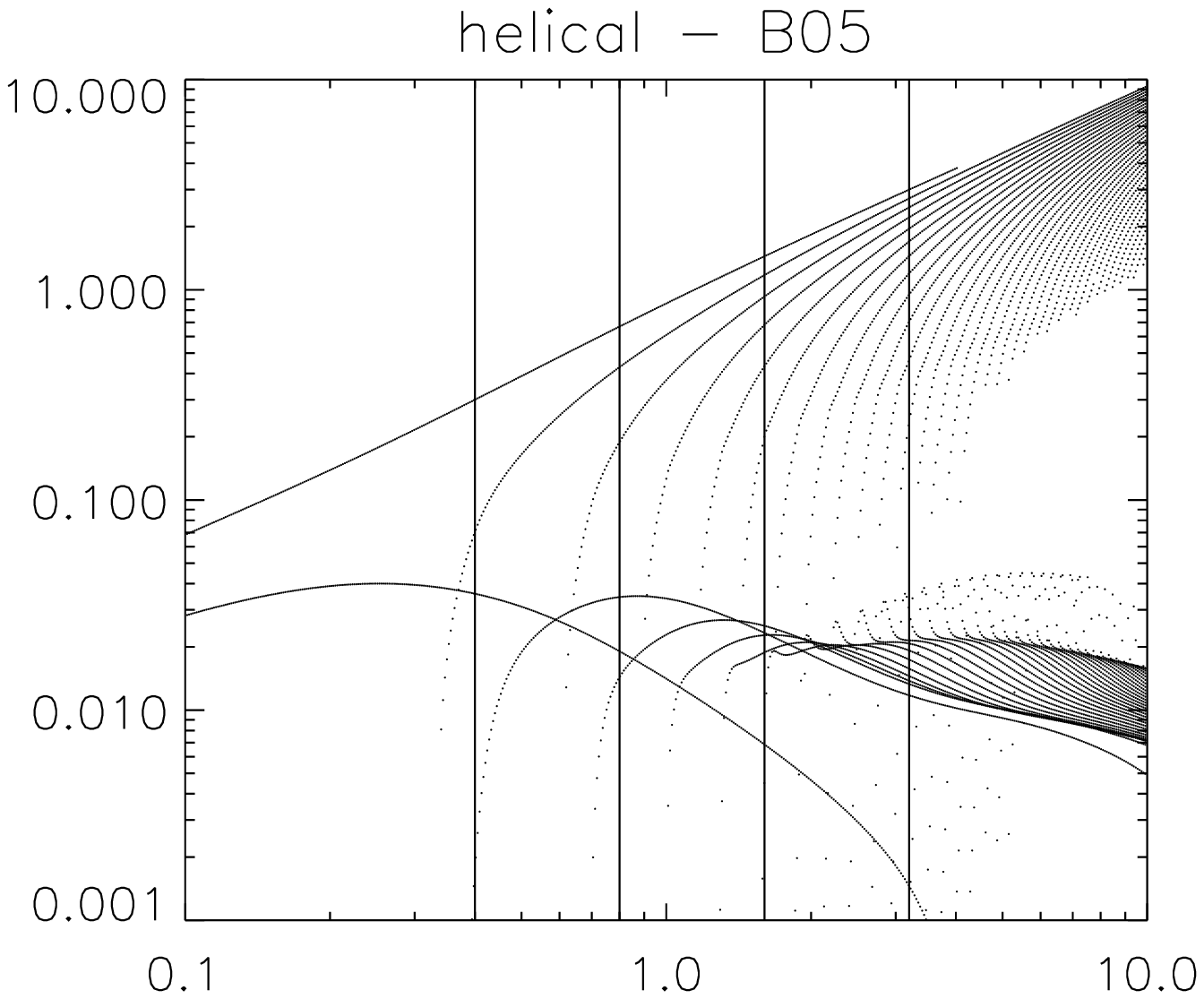}\quad
\includegraphics[clip,angle=0,width=0.3\textwidth]{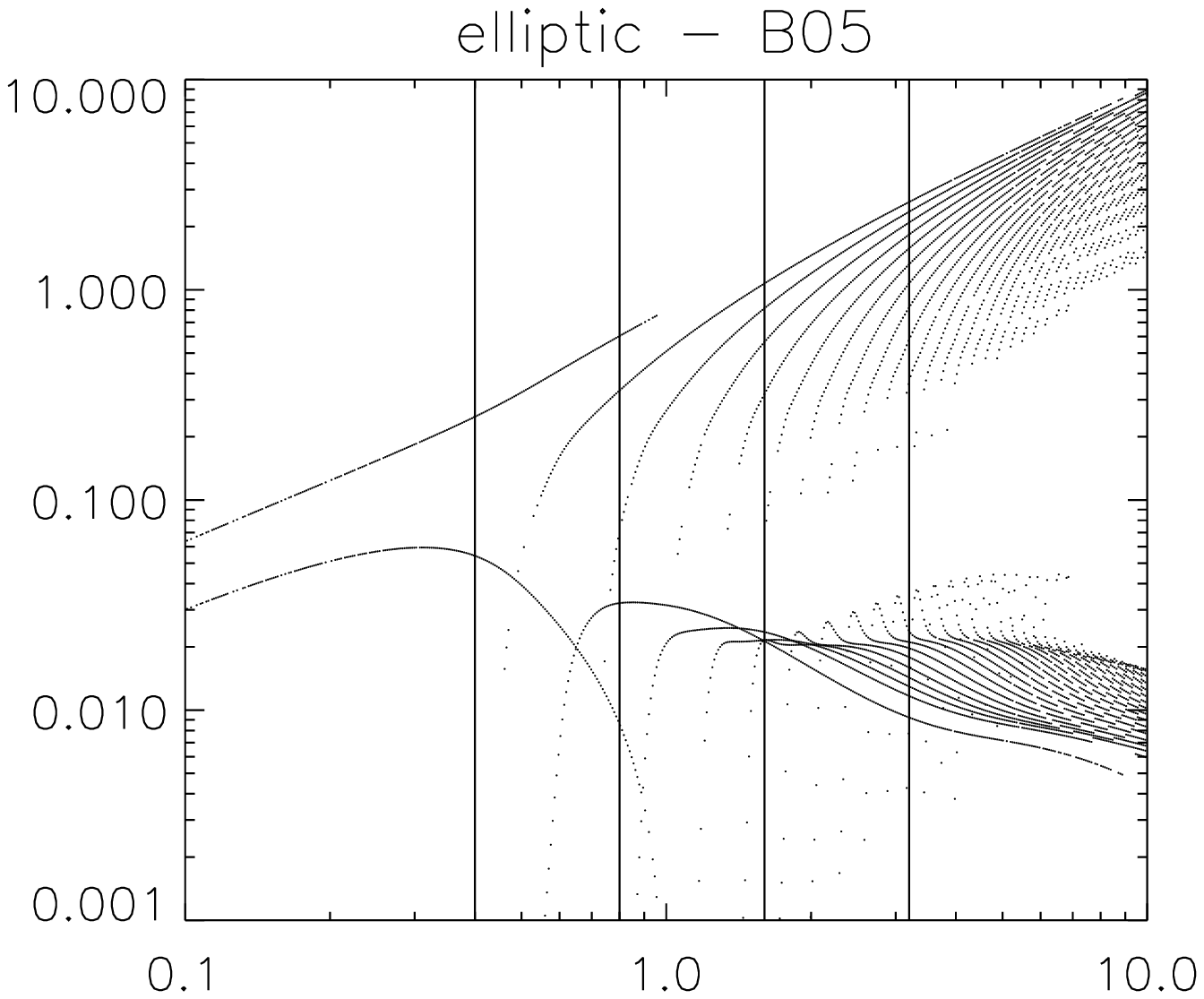}\quad
\includegraphics[clip,angle=0,width=0.3\textwidth]{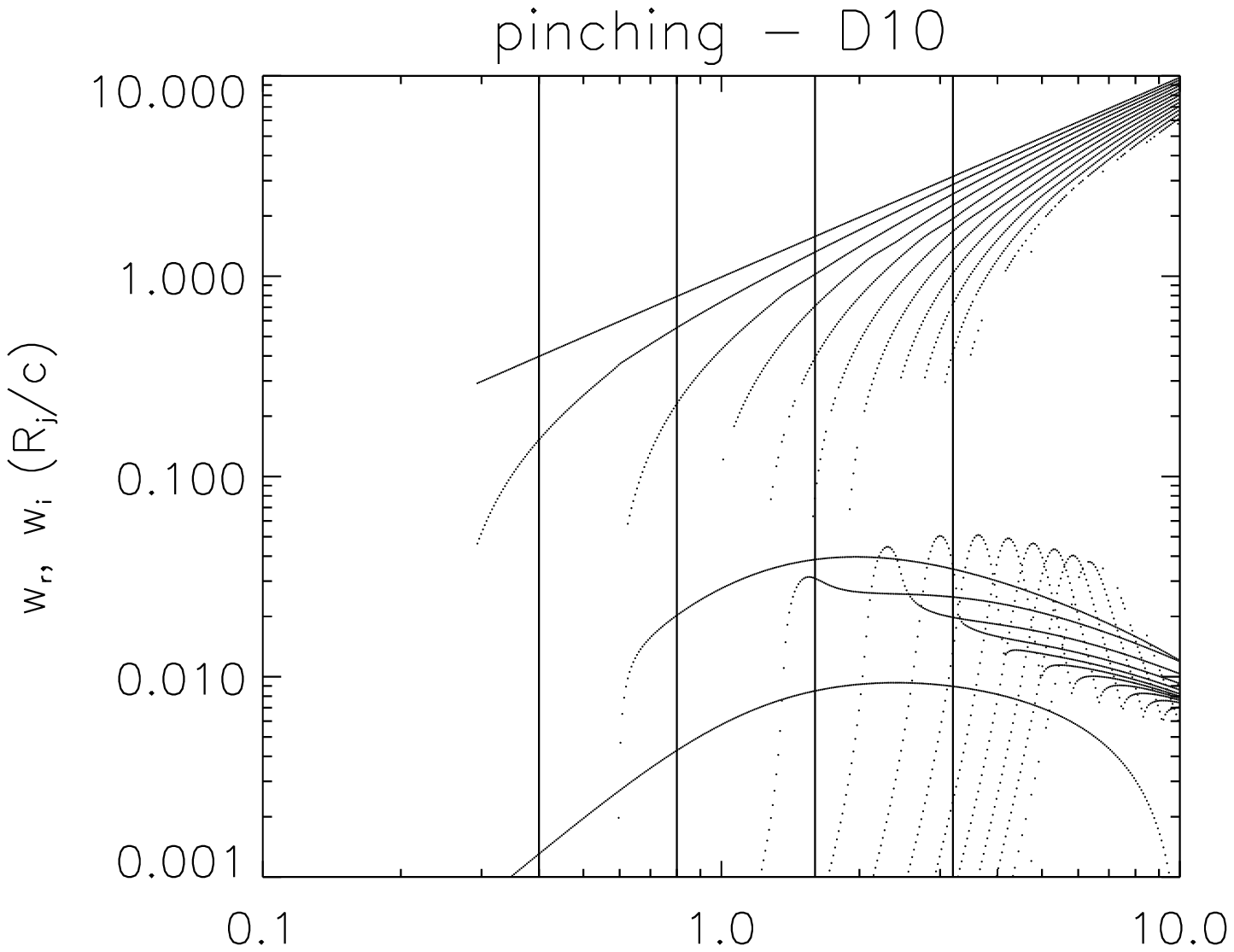}\quad
\includegraphics[clip,angle=0,width=0.3\textwidth]{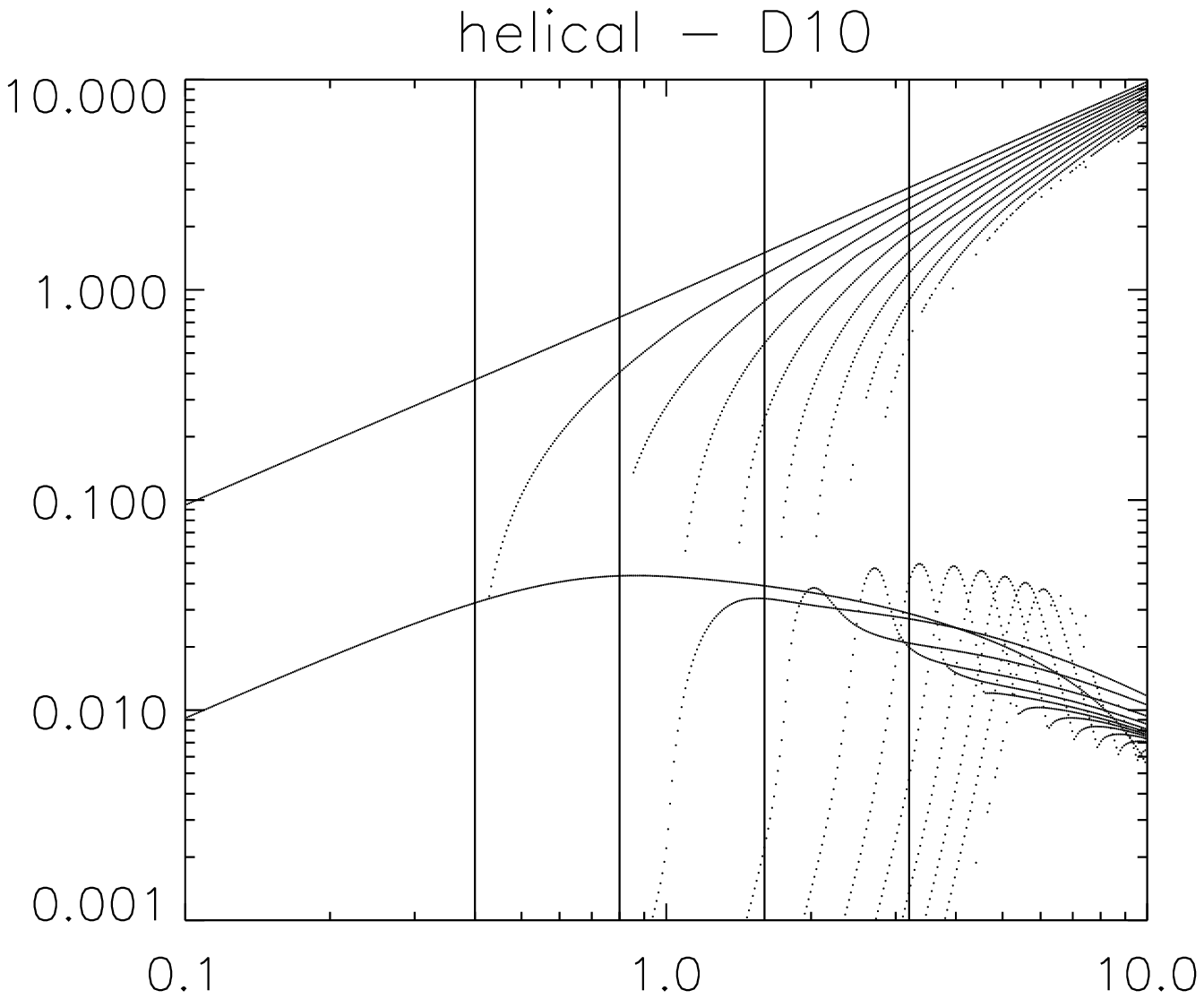}\quad
\includegraphics[clip,angle=0,width=0.3\textwidth]{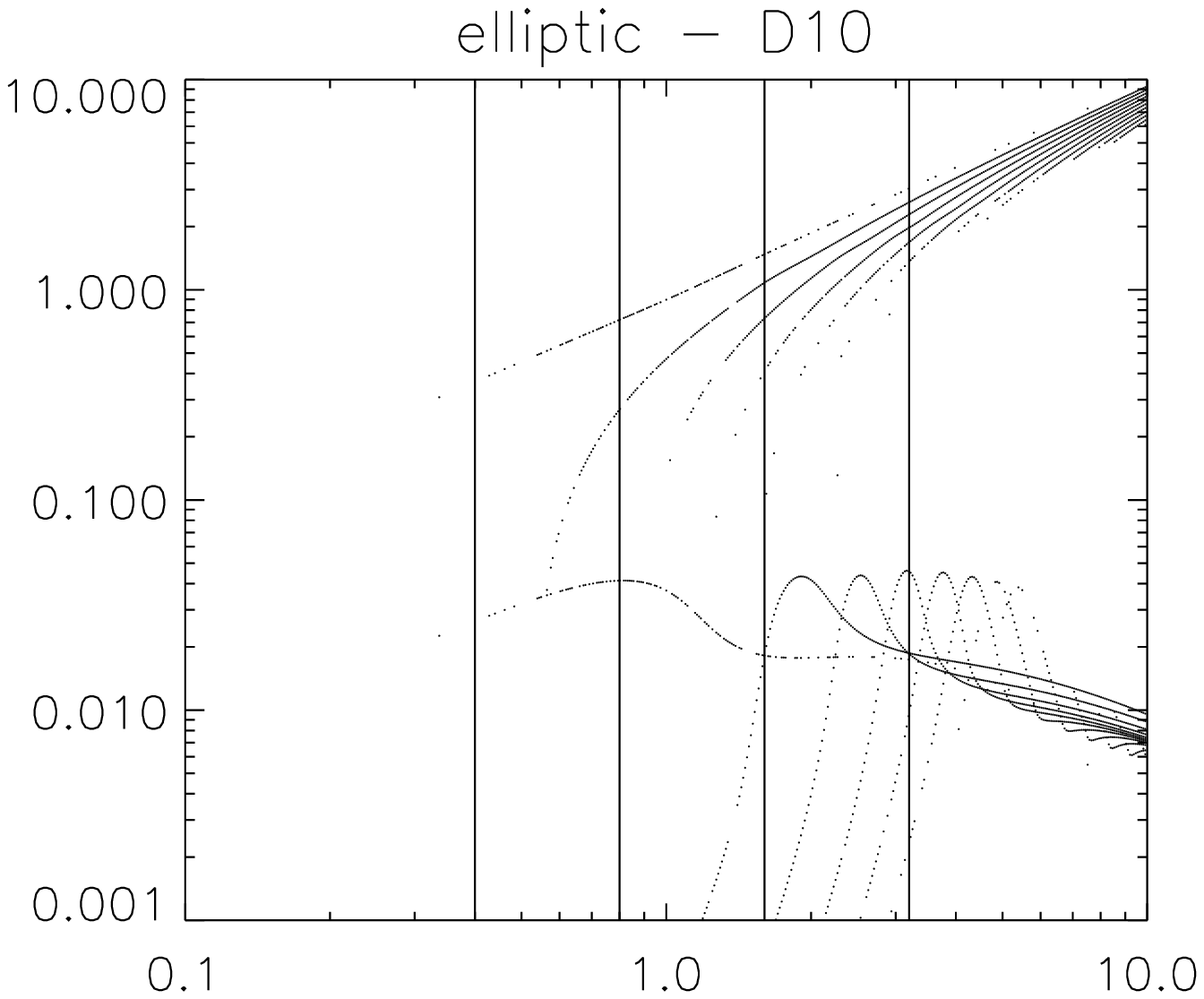}\quad
\includegraphics[clip,angle=0,width=0.3\textwidth]{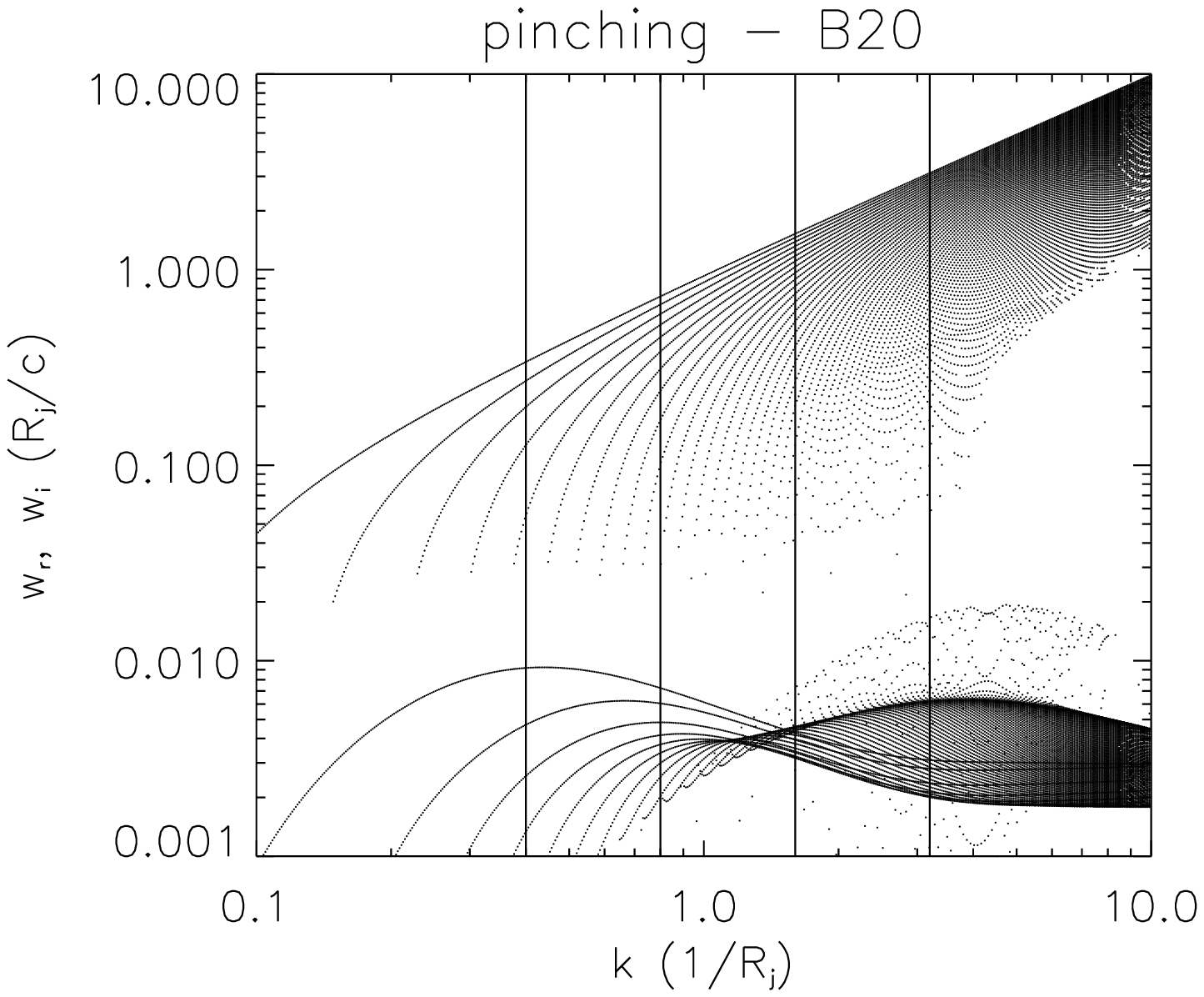}\quad
\includegraphics[clip,angle=0,width=0.3\textwidth]{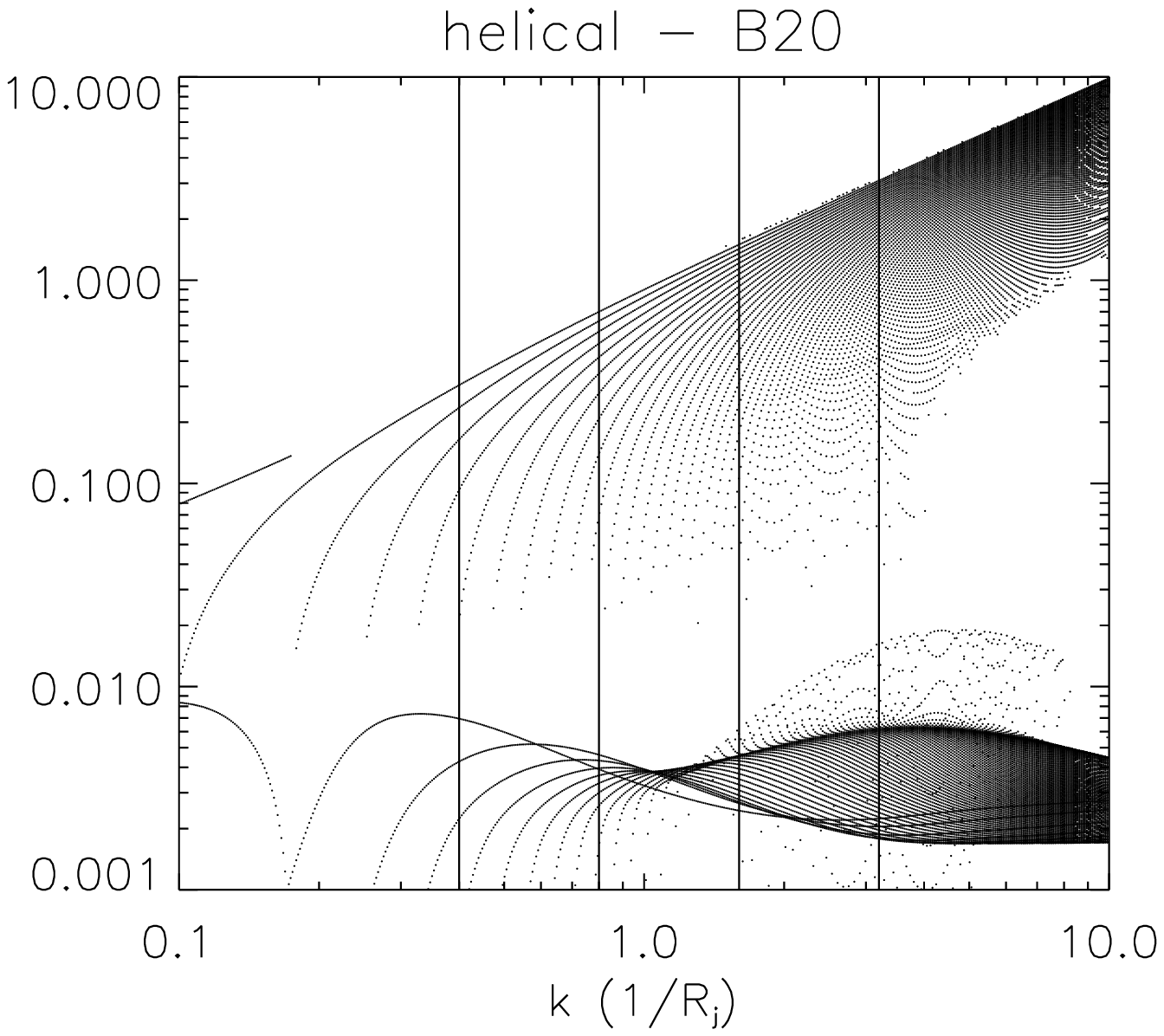}\quad
\includegraphics[clip,angle=0,width=0.3\textwidth]{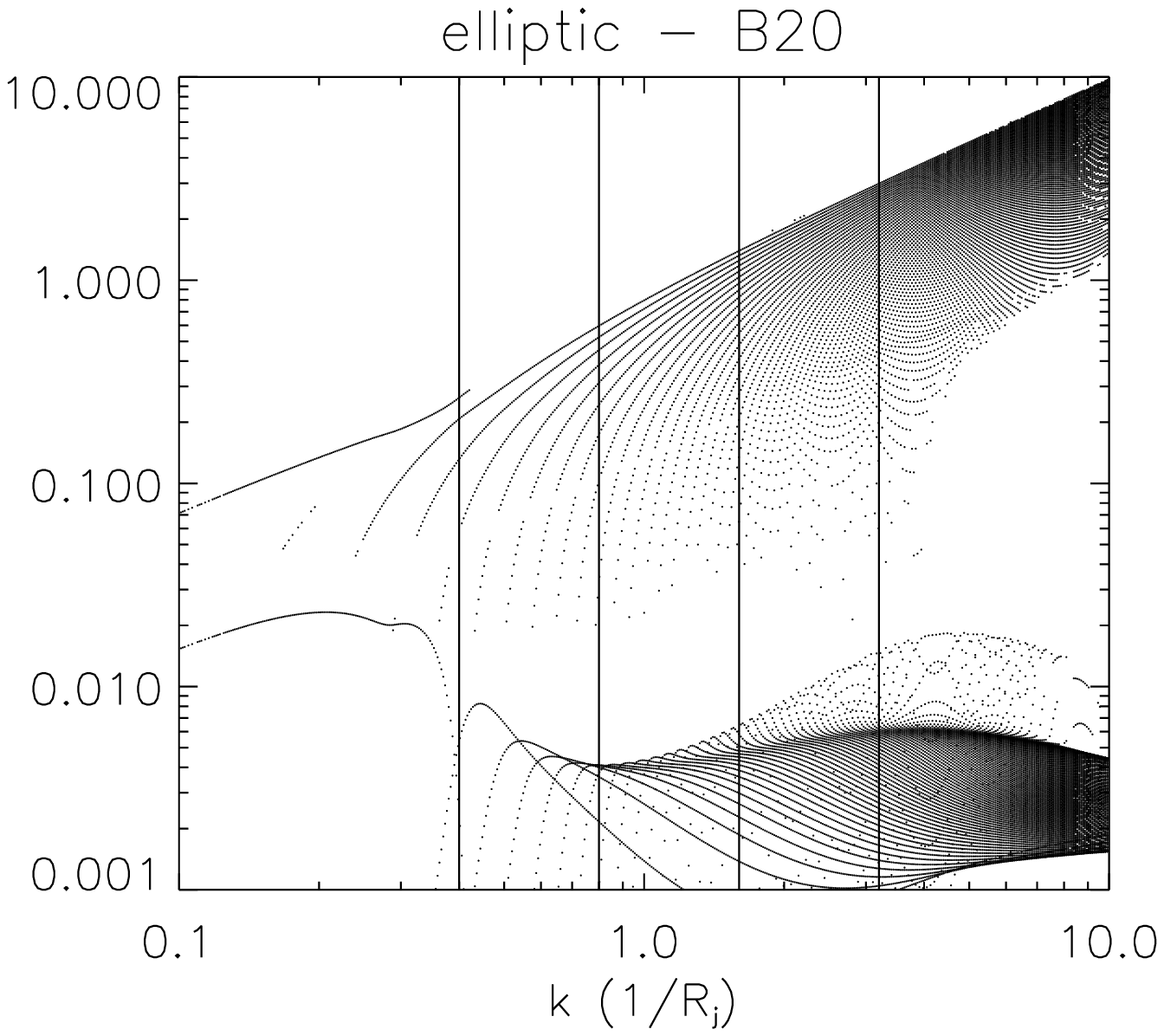}\quad
\caption{Solutions to the linear problem. Upper panels show the solutions for model B05, central panels show the solutions 
for model D10, and the bottom panels show the solutions for model B20. The left column displays the pinching modes, the central 
column shows the helical modes, and the right column shows the elliptic modes. In each of the panels, the upper lines represent 
the real part of the frequency, whereas the lower lines represent the imaginary part of the frequency, i.e., the growth rate. 
The surface modes are those with no zeros between the jet axis and its surface, whereas the body modes have one or more zeros 
in this region. They appear from left to right in the images.}
 \label{fig:linear}
 \end{figure*}

The solutions to the linear problem are shown in Fig.~\ref{fig:linear}. These solutions (for the wave-number and frequency 
of an instability) were obtained for the differential equation resulting from the linearization of the equations of 
relativistic hydrodynamics for a sheared jet in cylindrical coordinates \citep[e.g.,][for the vortex sheet case, and Paper~II 
for a sheared, slab geometry jet]{ha00}:
\begin{equation}\label{eq:lin}
 \frac{\partial^2 p}{\partial r^2} + F_1(r) \frac{\partial p}{\partial r} + F_2(r) p = 0,
\end{equation}
 with $r$ the radial coordinate, $p$ the perturbed pressure, 
\begin{equation}
 F_1(r) = \frac{1}{r} - 2 \frac{dv_{z,0}}{dr}\frac{\gamma_0^2 v_{z,0} (\omega - k_z v_{z,0}) + k_z } {\omega - k_z v_{z,0}} - \frac{d\rho'_0/dr}{\rho'_0 h_0}, 
\end{equation}
and
\begin{equation}
 F_2(r) = \frac{(\omega - k_z v_{z,0})^2 \gamma_0^2}{c_{sj,0}} - \frac{n}{r^2} + \left(\gamma_0^2 v_{z,0} (\omega - k_z v_{z,0}) + k_z \right) (\omega v_{z,0} - k_z),
\end{equation}
where $\gamma$ is the Lorentz factor, $v_z$ the axial velocity, $\omega$ the complex frequency, $k_z$ the complex wavenumber, 
$n$ the azimuthal wavenumber (0 for pinching modes, 1 for helical modes and 2 for elliptic modes), $\rho'$ the proper 
energy density, $h$ the specific enthalpy, $c_{sj}$ the sound speed of the jet and the subscript $0$ refers to the 
equilibrium values. The radial derivatives take a shear layer into account in the axial velocity and the energy density.   

Equation~(\ref{eq:lin}) was solved using the shooting method. This implies integration with a Runge-Kutta of variable 
step, from the axis to a given radius far from the jet boundary, for each pair of temptative values for $\omega$ and $k_z$. 
The boundary conditions on the jet axis are fixed in terms of the symmetry properties of the perturbation. The roots are 
found by fixing a real value of the wavenumber and looking for the roots (complex frequency) of the boundary condition that 
states that the amplitude of the wave must be zero far from the jet. The root-finder used with this purpose is the M\"uller 
method for equations of complex variable (check Paper~II for details). This process of calculation explains the point-to-point 
structure of the results shown. In addition, Bessel functions with complex arguments, which appear as solutions to 
eq.~(\ref{eq:lin}), are numerically computed. This is the first time that these kinds of results are systematically 
obtained for relativistic, sheared, cylindrical jets. 

The results shown in Fig.~\ref{fig:linear} reveal that the helical and elliptical surface modes dominate the growth rates 
at the longer wavelengths (low-order modes) for the three models. The plots show that the surface mode is damped at short 
wavelengths for colder jets (B05 and B20, top and bottom panels) when compared with the hotter jet (D10, central panels). 
The growth rates tend to decrease from B05 and D10 to B20, i.e., with the Lorentz factor and relativistic Mach number. The 
resonant modes dominate the solutions at the shorter wavelengths in all the models. Their growth rates are higher than those 
of the low-order modes. These results coincide with those obtained in Paper II for jets with slab geometry. 

We excited several modes for each model ($k\simeq 0.4,\,0.8,\,1.6,$ and $3.2\,R_j^{-1}$). In the case of B05, the 
pinching perturbations excite wavelengths surrounding the maximum growth rate of the first body mode, of the second body 
mode, and the resonant modes, whereas the helical and elliptical perturbations excite the surface, first body, second body, 
and resonant modes. In the case of D10 the pinching perturbations trigger the surface (at a very low growth rate), first 
and fourth body modes, and the helical and elliptical perturbations excite the surface and fourth and third body modes, 
respectively. Finally, in the case of B20, all the perturbations excite the first or second body modes and the resonant 
modes, as the surface mode is damped at the low values of the wave-number ($k<0.1\,R_j^{-1}$ for the helical surface 
mode and $k\simeq 0.4\,R_j^{-1}$ for the elliptical surface mode). We can thus anticipate from these results that the 
resonant modes may influence the growth of perturbations in all three cases, but will be more important in the case of 
pinching and helical modes in B20, due to their higher relative growth rates.

 \begin{figure*}[!t]
    \centering
 \includegraphics[clip,angle=0,width=0.45\textwidth]{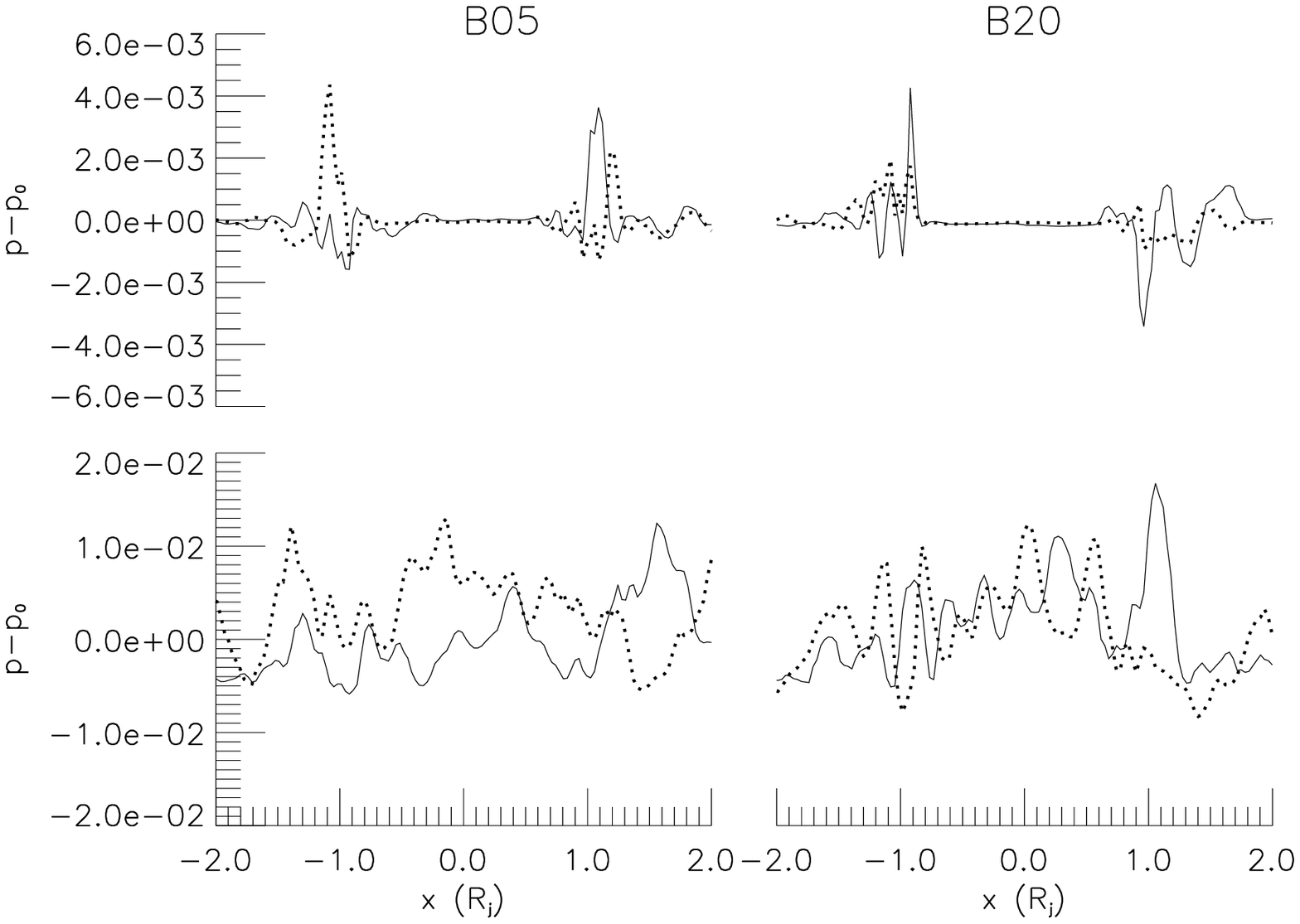}\quad
\includegraphics[clip,angle=0,width=0.45\textwidth]{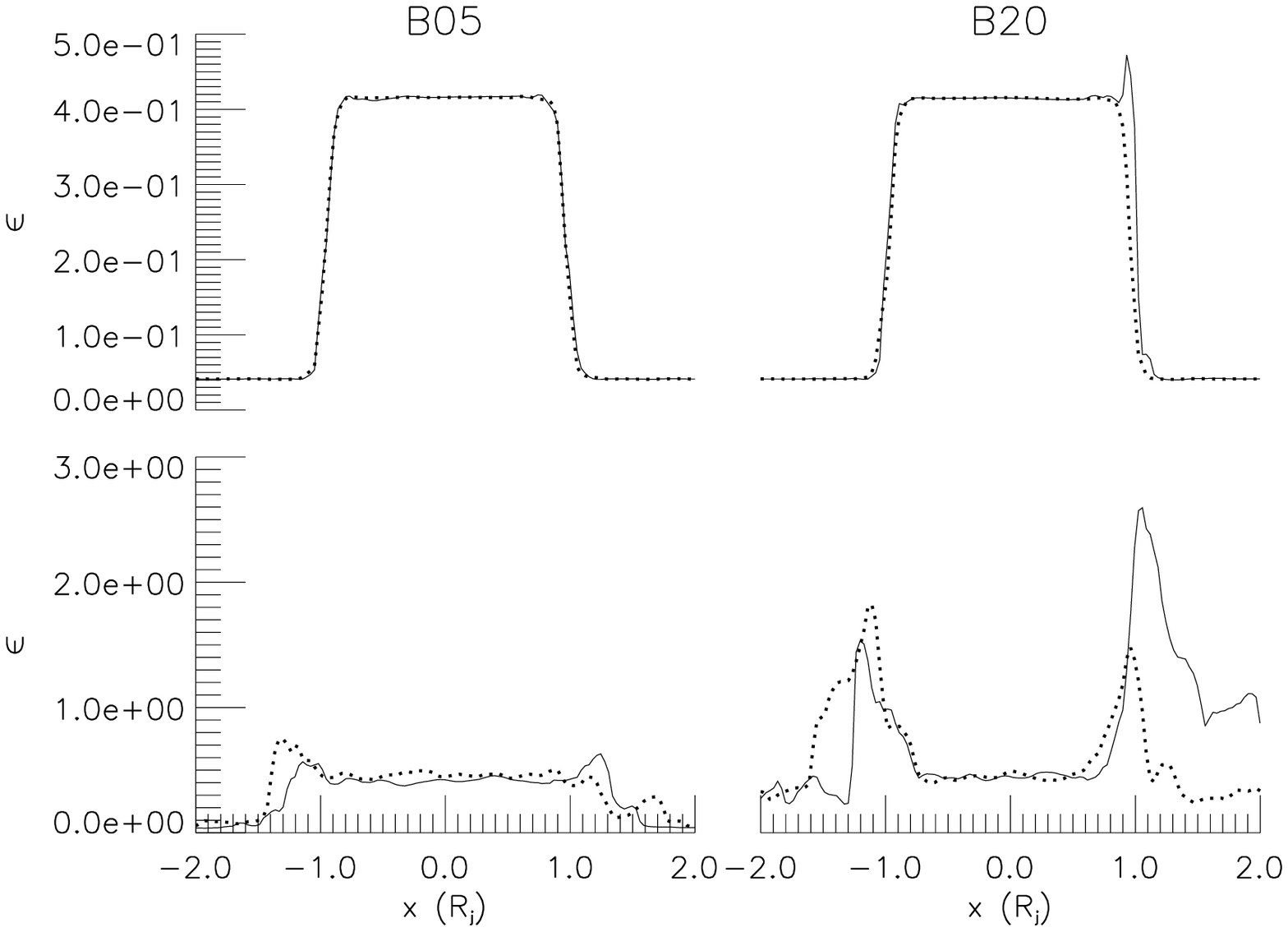}
 \caption{Left panels: Pressure perturbation across the jet axis for models B05 (left panels) and B20 (right panels) at 
two different times ($t_1$ and $t_2$ in Fig.~\ref{fig:mom} for B05 and $t_1$ and $t_3$ for B20), within the linear 
regime (upper row) and close to the end of the linear regime (lower row). Arrows in Fig.~\ref{fig:mom} indicate these 
times. Right panels: Specific internal energy across the jet axis for models B05 (left panels) and B20 (right panels) at 
the same times. The two lines (solid and dotted) stand for two different positions along the jet axis, at half the grid-size 
distance to each other, where the cuts are done. Note the change in scale between the upper and lower plots.}
 \label{fig:resonant}
 \end{figure*}

During the linear evolution, the resonant modes are excited either directly or as harmonics of the longer wavelengths and 
may grow faster than the latter, as explained in the previous paragraph. In Paper II, the authors showed that these modes 
grow faster in the shear layer between the jet and the ambient. Thus if the resonant modes dominate the growth, the largest 
mode amplitudes are expected towards the shear layer. The upper panels in Fig.~\ref{fig:resonant} show the typical profiles 
produced by the growth of resonant modes (see Figs. 4 and 9 in Paper II) for models B05 (left) and B20 (right). The radial 
profile of the perturbations in these jets is clearly dominated by the resonant modes. However, this does not imply that 
they have the highest amplitude in the whole volume of the jet, as revealed by a Fourier analysis along the jet (not shown 
here), which shows that they are in competition with the longer wavelengths, mainly in the case of B05. Axial profiles of 
the pressure perturbation (not shown) taken at $r=0.5\,R_j$ reveal long antisymmetric structures along the jet for model 
B05, together with the short scale oscillations, whereas in the case of B20, the long structures appear to be mainly 
symmetric (due to the high growth rate of the elliptical modes) and short scale oscillations of comparable amplitude 
are also visible.

The lower panels in Fig.~\ref{fig:resonant} show the same variables at a later time, close to the end of the 
linear regime \citep[as defined in][]{pe+04a}. The evolution of the pressure perturbation in time is very similar, 
in both cases, to that shown in Fig. 9 of Paper II, with the larger amplitudes initially showing up in the sheared region. 
We show, in the right panels of Fig.~\ref{fig:resonant}, two cuts of specific internal energy across the jet for both models, 
at the same times as in the left panels. Both the upper (in the linear regime) and lower (about the end of the linear regime) 
plots reveal that the resonant modes convert more energy into heat in the shear layer in the case of model B20. Thus, as 
expected from the linear solutions (see Fig.~\ref{fig:linear}), these modes influence the dynamics of this jet more.

\subsection{Nonlinear regime}

\begin{figure}[!h]
    \centering
 \includegraphics[clip,angle=0,width=\columnwidth]{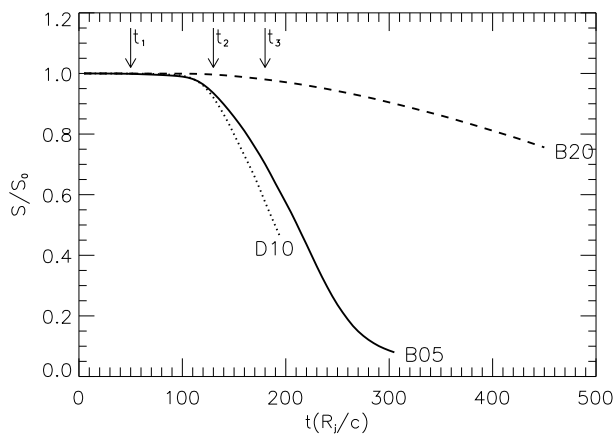}
 \caption{Evolution of axial momentum in the jets with time. The solid line stands for model B05, the dotted line for 
model D10 and the dashed line for model B20. Indicated with arrows, the times at which the cuts in Fig.~\ref{fig:resonant} 
were taken: $t_1$ is the time of first cut for both models, $t_2$ the time of the second cut for B05, and $t_3$ the 
time of the second cut for B20.}
 \label{fig:mom}
 \end{figure}

In previous works the stability of jets in the nonlinear regime was characterized in terms of mixing and 
transfer of linear momentum to the ambient medium \citep[][and Paper I]{pe+04b}. Following these criteria, we discuss here 
the stability of the simulated jets. The axial momentum in the simulated jets is displayed versus time in Fig.~\ref{fig:mom}, 
normalized to their value at $t=0$. The longitudinal momentum in the jet is computed by multiplying the actual axial momentum 
in each cell by the jet mass fraction. During the linear regime, we observe no transfer of linear momentum from the jet to the 
ambient. This regime is shorter in these simulations than in the 2D ones because the initial amplitudes of the 
perturbations have been set to be larger with the aim of reducing the computational times. In the nonlinear regime, the plots 
show a clear trend following the results reported in Paper I: model B05 loses the largest amount of axial momentum and model 
B20 keeps a large part of it at the end of the simulation. The latter was kept running longer than models B05 and D10 to check 
that this trend is valid during long times in the nonlinear regime. Model D10 was stopped at an earlier stage than B05 because the 
time-step became progressively shorter during the nonlinear regime and the simulation demanded too much time to be continued from 
this point.    

\begin{figure*}
    \centering
 \includegraphics[clip,angle=0,width=0.45\textwidth]{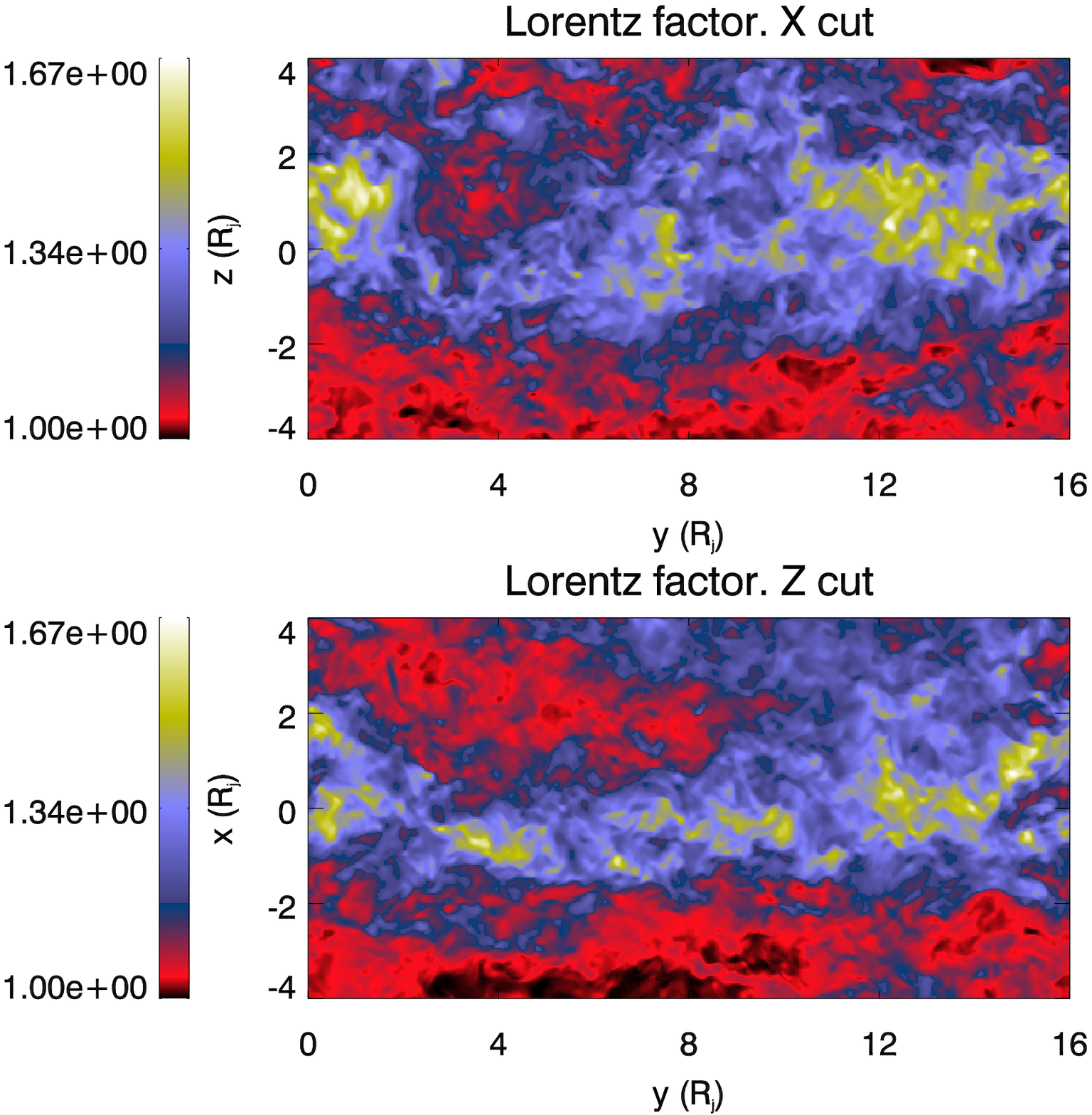}
\quad
\includegraphics[clip,angle=0,width=0.45\textwidth]{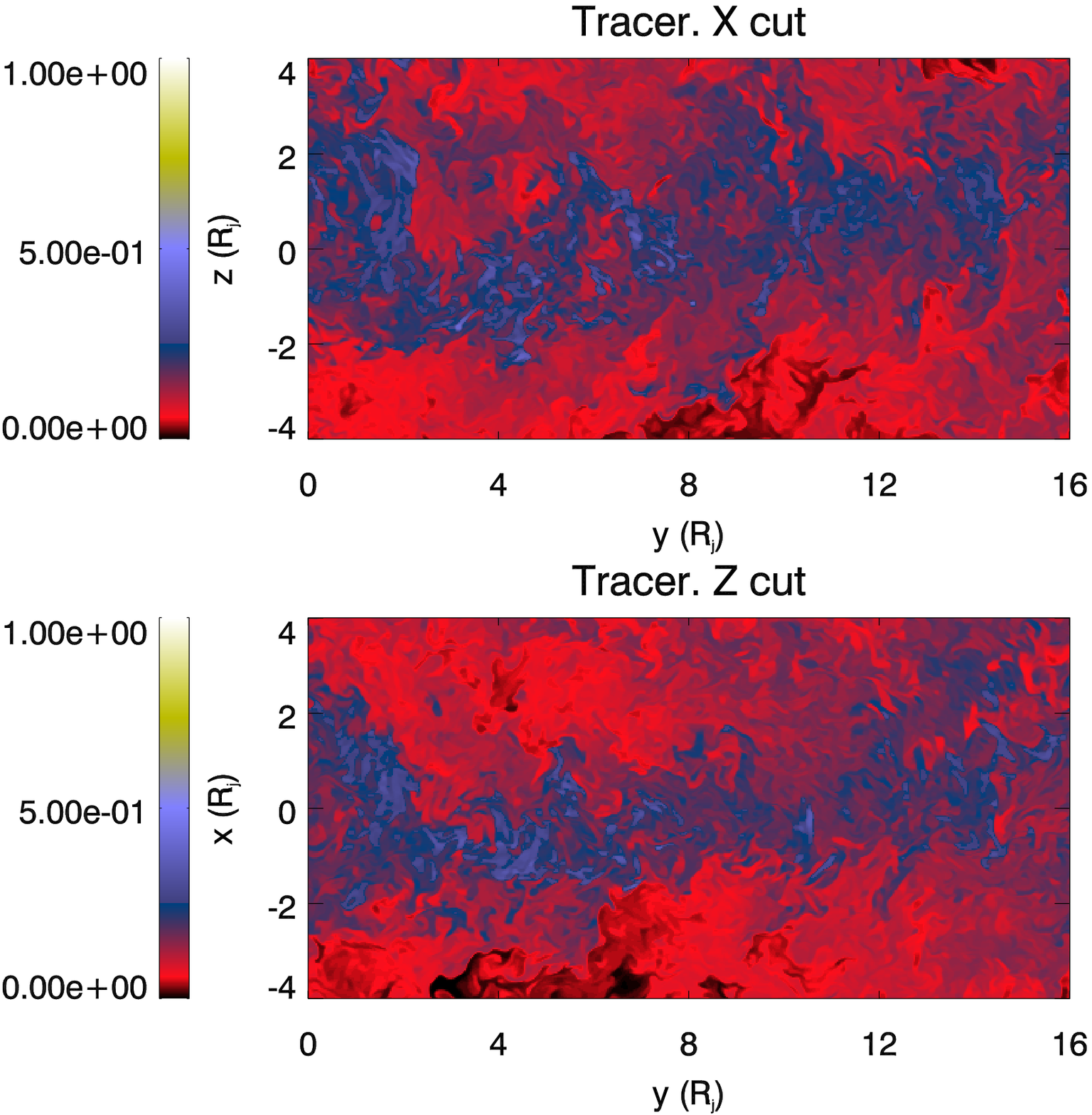}
 \caption{Axial cut of Lorentz factor (left) and jet mass fraction (tracer, right), along the central planes for 
model B05 at the end of the simulations. The tracer has a value of 1 for the original jet material, 0 for the ambient 
medium material and values between 0 and 1 in those cells where mixing is present.}
 \label{fig:b05}
 \end{figure*}

\begin{figure*}
    \centering
 \includegraphics[clip,angle=0,width=0.45\textwidth]{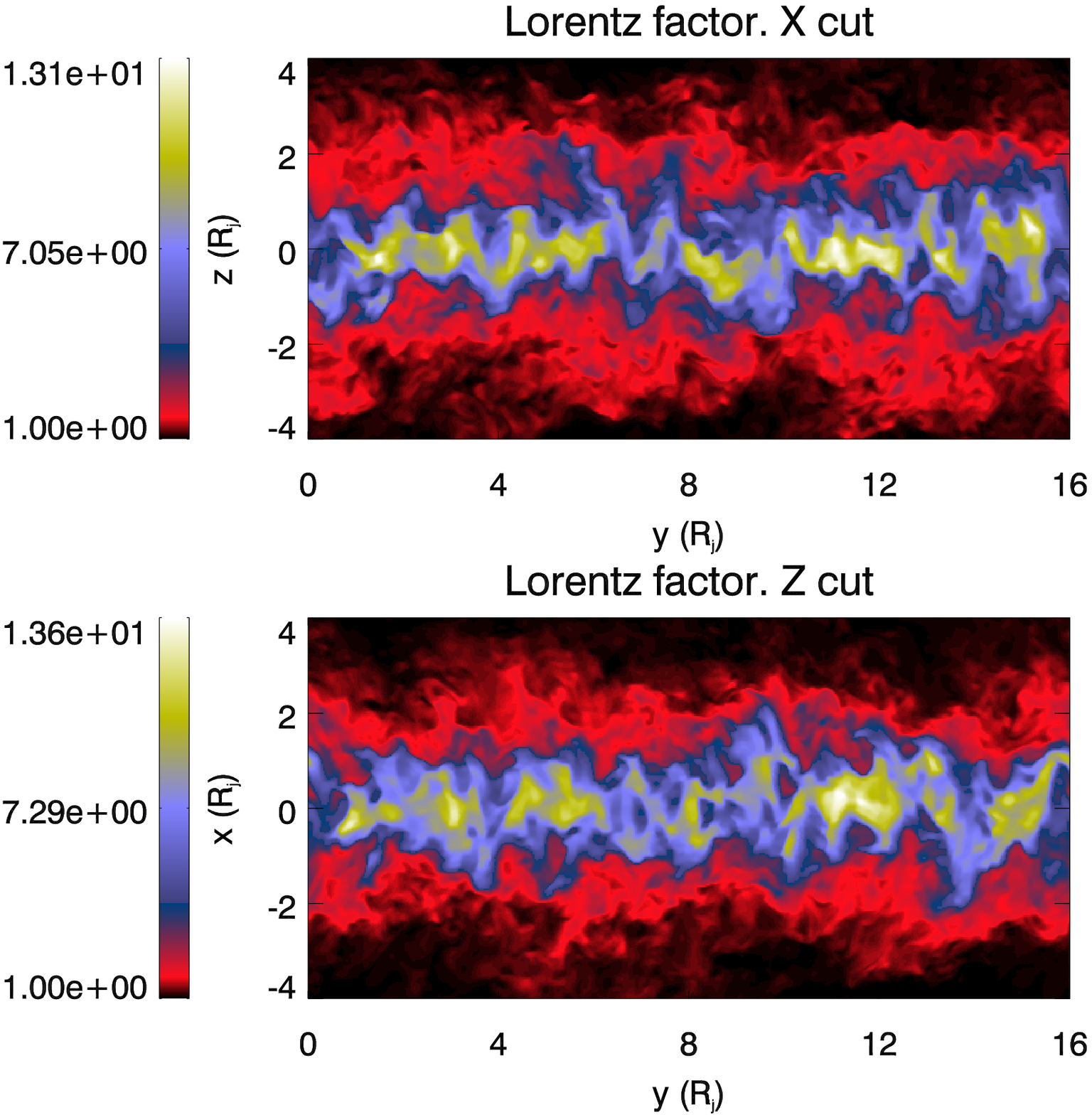}
\quad
\includegraphics[clip,angle=0,width=0.45\textwidth]{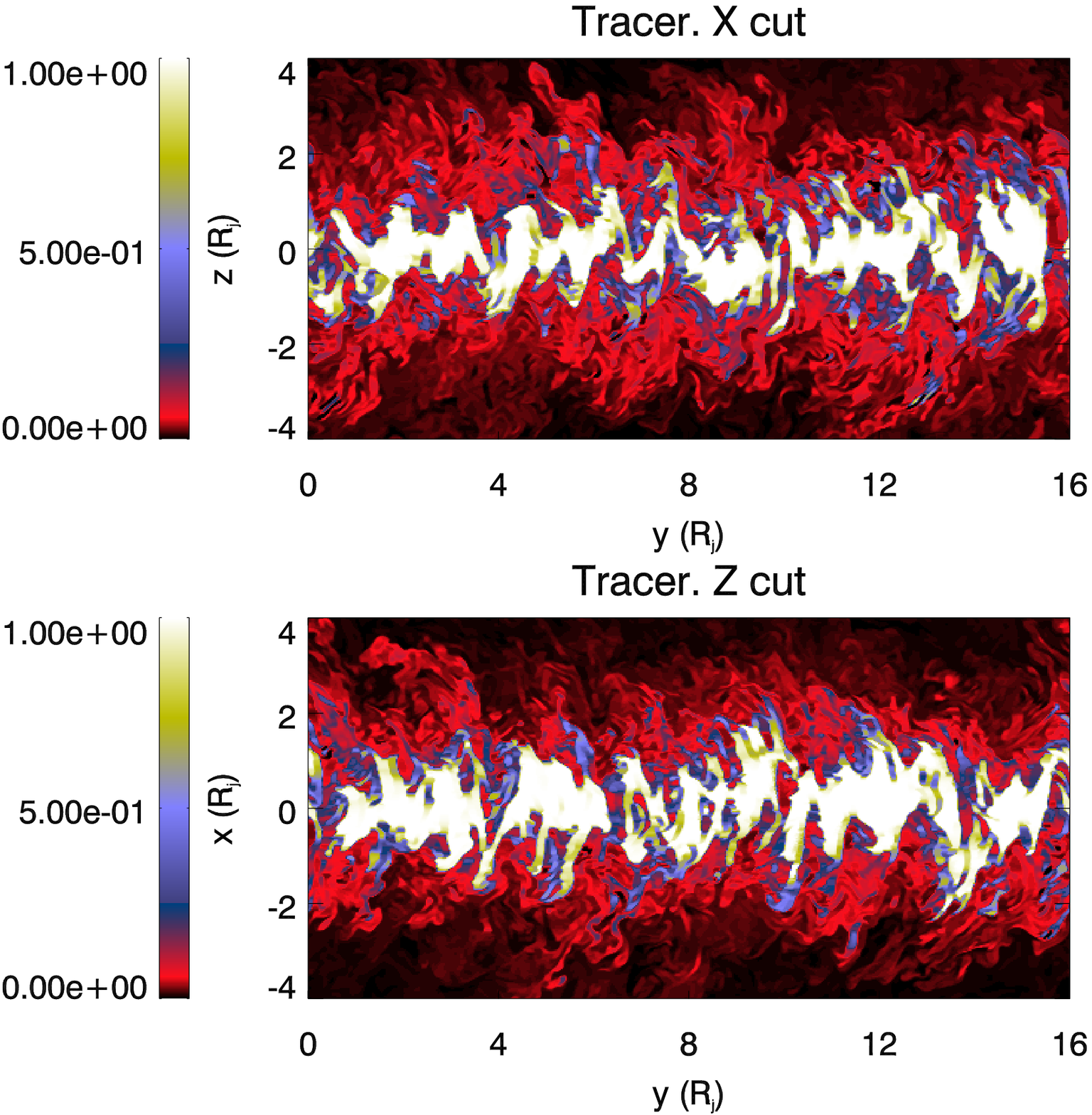}
 \caption{Axial cut of Lorentz factor (left) and jet mass fraction (tracer, right), along the central planes for model 
D10 at the end of the simulations.}
 \label{fig:d10}
 \end{figure*}

\begin{figure*}
    \centering
 \includegraphics[clip,angle=0,width=0.45\textwidth]{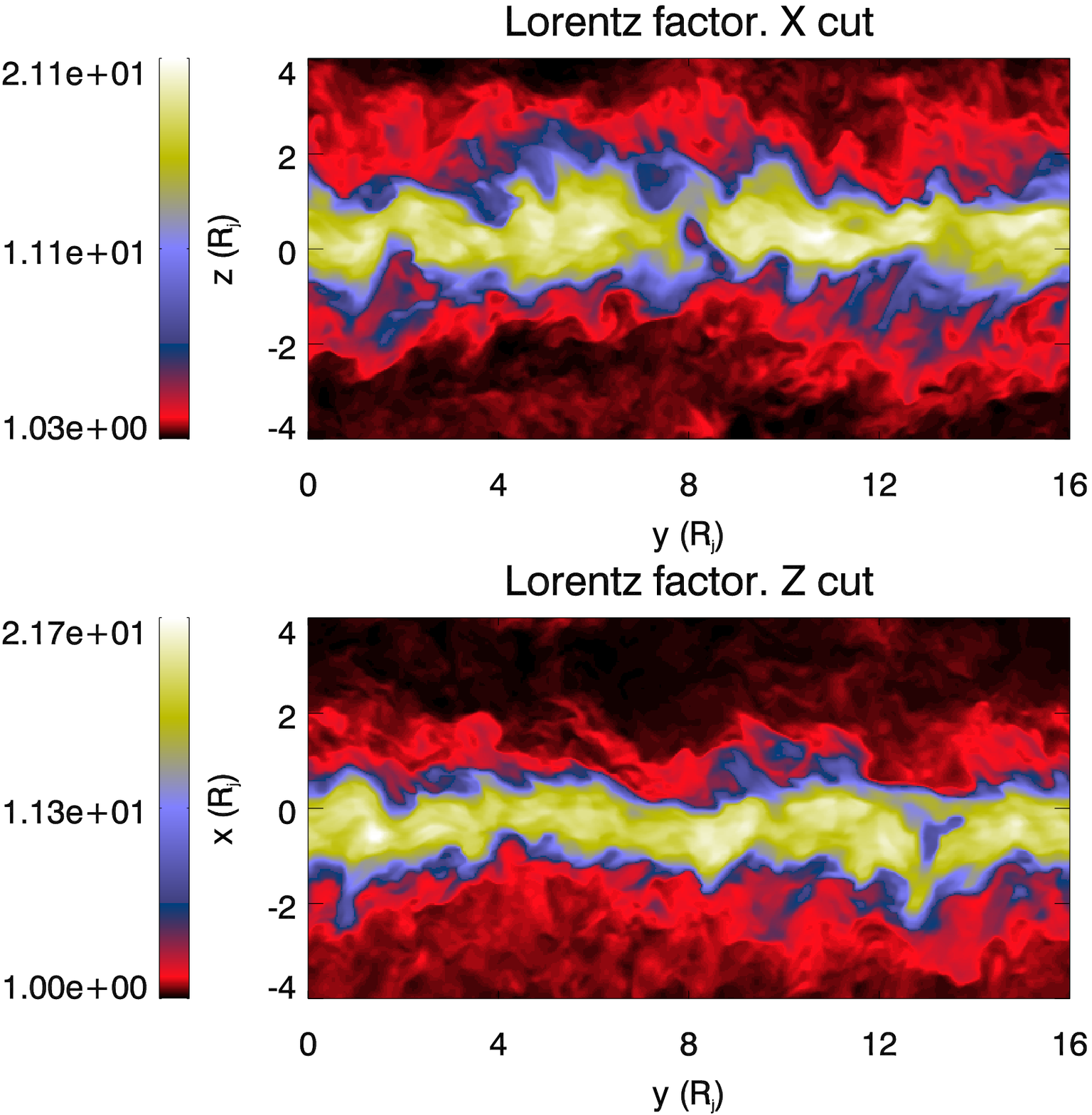}
\quad
 \includegraphics[clip,angle=0,width=0.45\textwidth]{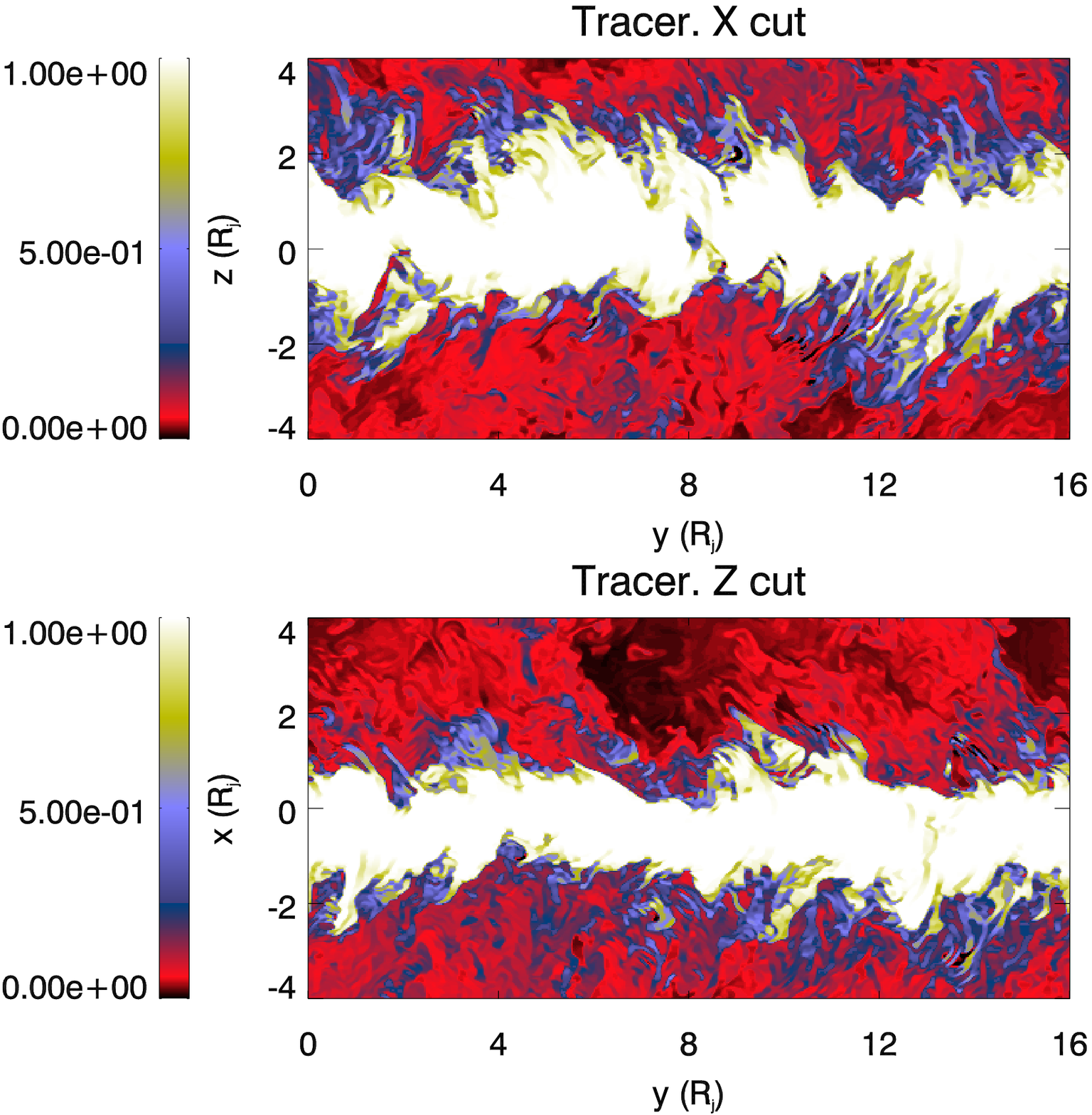}
 \caption{Axial cut of Lorentz factor (left) and jet mass fraction (tracer, right), along the central planes for model 
B20 at the end of the simulations.}
 \label{fig:b20}
 \end{figure*}

Figures~\ref{fig:b05} (model B05), \ref{fig:d10} (D10), and \ref{fig:b20} (B20) display axial cuts at the final snapshot in 
Lorentz factor and tracer (jet mass fraction) for the three simulations. The maximum Lorentz factor in the maps of models 
B05 is $\gamma = 1.67$, in the maps of model D10 is $\gamma = 13.1-13.6$, and $\gamma = 21.1-21.7$ for model B20. Regarding 
the tracer, white indicates original, unmixed jet material. From these maps, we see that B05 is completely mixed and D10 is 
strongly entrained, although both models keep a central region with the higher values of this variable, whereas model B20 
keeps a central spine with little or no mixing. 

The maps show the effect of the different modes in the structure of the flow and a gradient in the stability of the jets from 
B05 and D10 to B20, as could be expected from Fig.~\ref{fig:mom}. Although the mixture of growing modes makes it difficult 
to identify the individual ones, these images allow us to see which wavelenghts dominate the nonlinear regime. Model B05 
shows a helical structure with the size of the box that can be attached to the surface mode (see Fig.~\ref{fig:linear}), 
but also small-scale efficient mixing structures resulting from the growth of the short-wavelength resonant modes. The 
Fourier transform shows competition among different harmonics, mainly the fundamental mode of the box, followed 
in amplitude by the first harmonic and a mixture of higher order harmonics, which coincides with the predictions from 
the linear regime. The rapid disruption of this jet is the consequence of the distortion of the jet surface by the longer 
wavelengths.

The central spine of model D10 shows a short wavelength ($1-2\,R_j$), helical structure, which could be identified with 
the fourth body mode (see Fig.~\ref{fig:linear}), and, on the same scales, expansions and contractions of the jet, which 
could come either from the fourth pinching body mode or from the third elliptical body mode. These structures are surrounded 
by a layer where small-scale mixing occurs. In this region, an $8\,R_j$ helical pattern can be seen. This result clearly 
shows that different wavelengths may show up at different jet radii, generating a radial profile, as already shown in 
\cite{pe+06}. The Fourier transform indicates that the highest amplitudes correspond to the fundamental mode and first 
harmonic at the end of the linear regime, followed by high-order harmonics, which keep growing slowly after the end of 
the linear regime until saturation of the growth of the radial velocity and domination of the nonlinear regime in the 
long term. The growth rates shown in Fig.~\ref{fig:linear} agrees with the importance of the third to sixth body modes during 
the evolution. However, in contrast to the colder models, the linear solution does not show small-scale 
resonances for high-order body modes, but the resonances occur for those low-order body modes, which may be the reason 
for the final disruption of the jet. The structure of jet D10, with maxima and minima in the Lorentz factor along the 
axial directions will make shock waves arise naturally, which may mix the jet further and eventually disrupt it.

Model B20 presents a more collimated structure and shows a combination of expansions and contractions plus a helical 
pattern with the wavelength of the box, corresponding to the first body helical mode. As in model D10, there is a thick 
mixing layer surrounding the jet core. The Fourier transform also shows that the fundamental mode of the box (corresponding 
to the helical first body mode) dominates during the linear regime, but also the fast growth of the higher harmonics, which, 
like before, surpass the amplitude of the long wavelength during the time interval between the end of the linear regime and 
the saturation of the radial velocity. This competition between the longest wavelength and the shorter ones can be deduced 
from Fig.~\ref{fig:linear}. In the case of model B20, the structure at the end of the simulation points towards further mixing, 
although this jet has a higher Lorentz factor and a more uniform structure than D10 at the end of the simulation.

\begin{figure*}
    \centering
 \includegraphics[clip,angle=0,width=\textwidth]{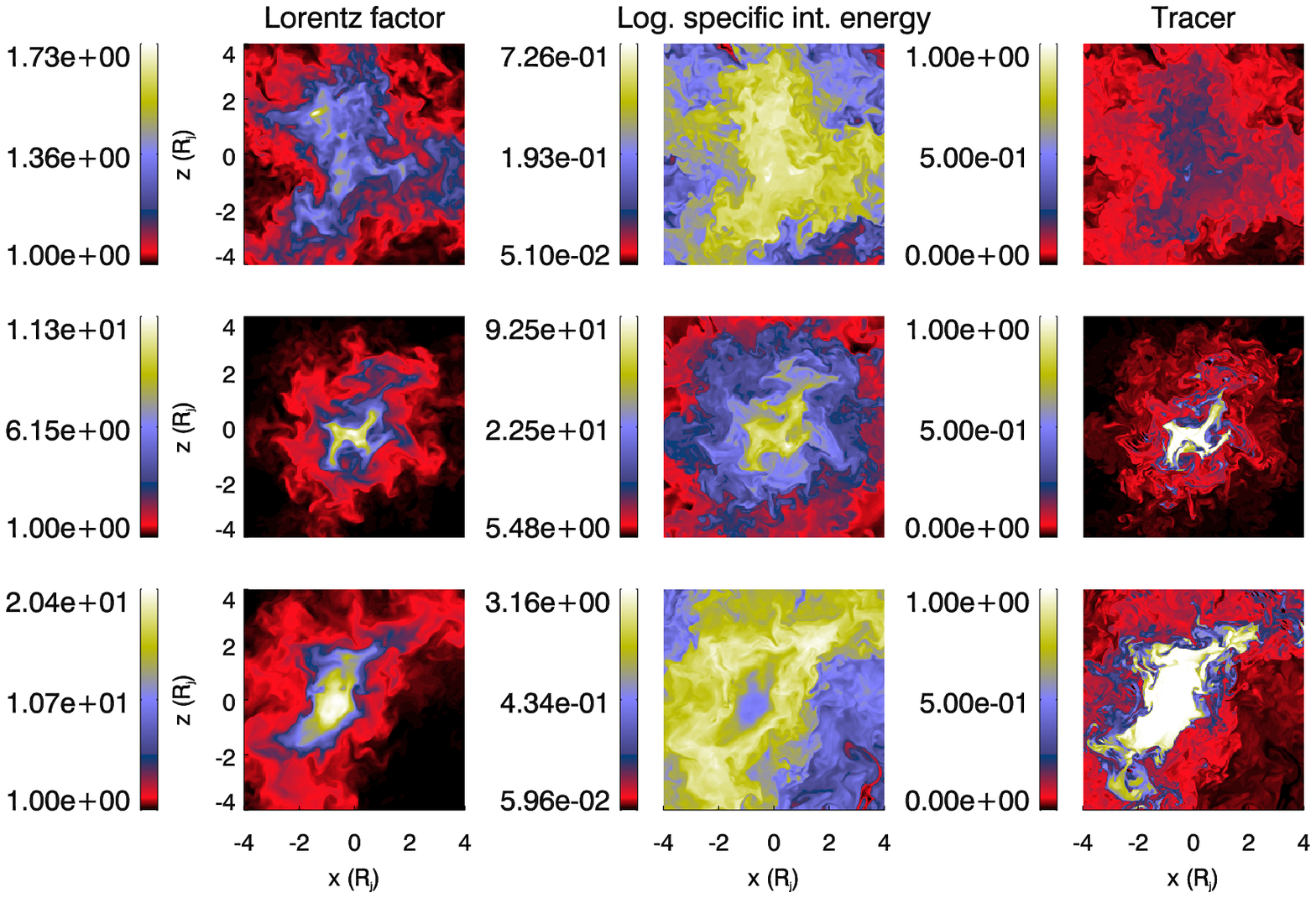}
 \caption{Transversal cut of Lorentz factor (left), logarithm of specific internal energy ($c^2$, centre) and jet 
mass fraction (tracer, right), at $y=8\,R_j$ for model B05 (top panels), D10 (middle panels), and B20 (bottom panels) 
at the end of the simulations.}
 \label{fig:prrof}
 \end{figure*}

Figure~\ref{fig:prrof} shows transversal cuts at half way through the grid at the end of the simulation. The figure shows 
the final distorted structure and turbulent mixing in all three models. The top panels show that model B05 is completely 
mixed (see the tracer in the right panel), whereas D10 and B20 still keep a fast, unmixed spine, which is larger in the 
latter. B20 shows a hot shear layer including the slower, outer regions of this unmixed core, contrary to D10 and B05, where 
the temperature decreases outwards. The centroid of model B20 is displaced left and down from the centre of coordinates, 
indicating helical motion in competition with the elliptical modes, responsible for the oblique elongation of the jet section. 
There are hints of a similar deformation in model D10, but no significant displacement of the centre of the jet is detected in 
this case. This may be because the dominating helical structure is given by a short wavelength mode, which only 
generates small displacements \citep{ha00}. Also model B05 shows an elongation (vertical in this cut), but any hints of regular 
structure have been smeared out by the mixing and disruption of the original flow. It is also difficult to assess whether the 
axis of the jet is displaced by helical modes; this can be better observed in the axial cuts.

\begin{figure*}
    \centering
 \includegraphics[clip,angle=0,width=\textwidth]{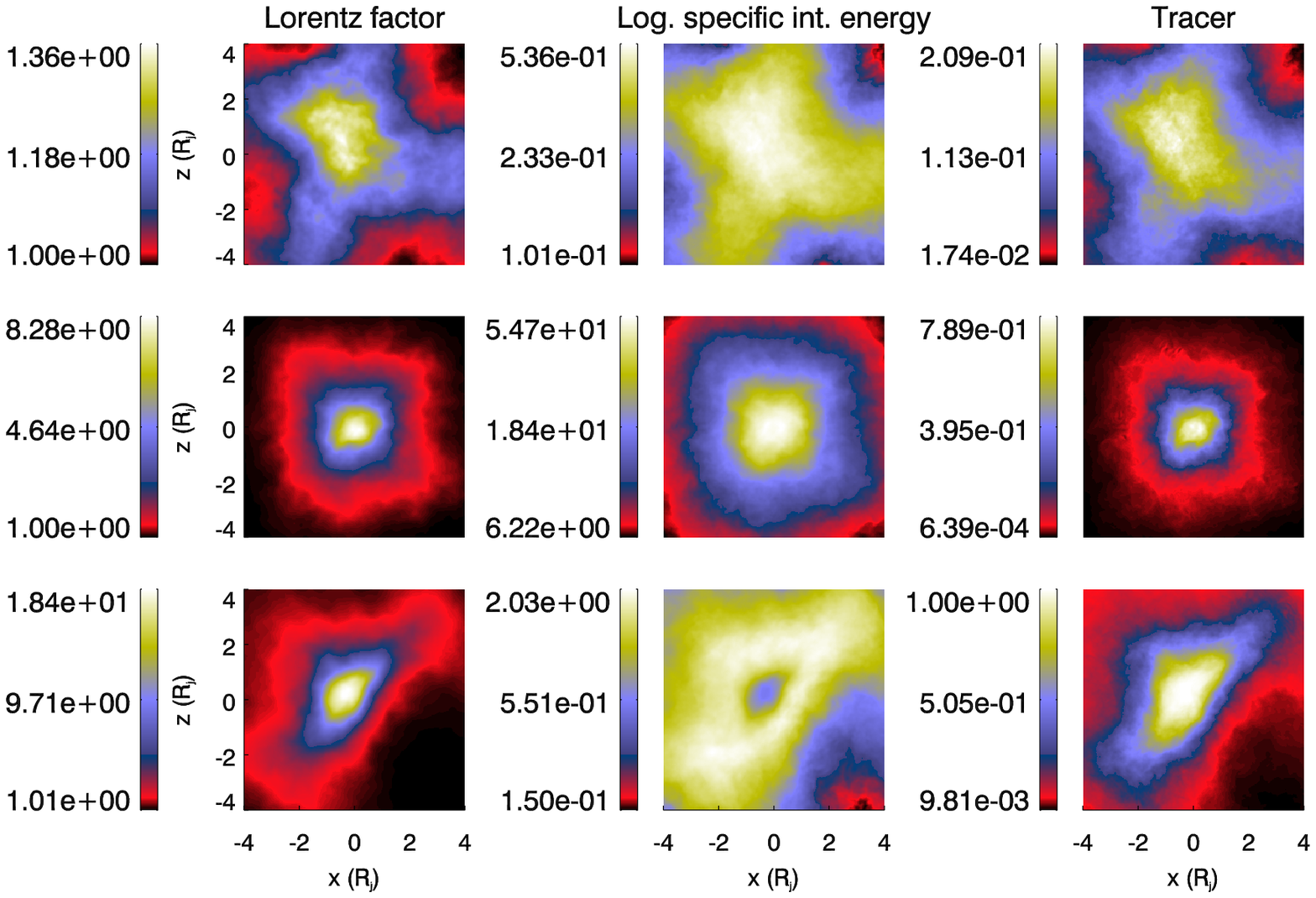}
 \caption{Mean values of Lorentz factor (left), logarithm of specific internal energy ($c^2$, centre), 
and jet mass fraction (tracer, right), averaged along the jet axis for model B05 (top panels), D10 (middle panels), 
and B20 (bottom panels) at the end of the simulations. Scales on the different panels are fixed to span the interval 
between the local minima and maxima.}
 \label{fig:profs}
 \end{figure*}

To have a global view of the parameters along the jet axis, we show in Fig.~\ref{fig:profs} mean axial values 
of different physical parameters for the three models at the end of each simulation. In these panels, the displacement of 
the axis due to the helical modes is promediated and thus we always find the centre of the jets around their original 
position. The left panels display the Lorentz factor and show barely relativistic motion for model B05, with maximum 
Lorentz factor $\gamma=1.36$, as compared to models D10 ($\gamma=8.28$) and B20 ($\gamma=18.4$). The central panels show 
the logarithm of the specific internal energy and show the same behaviour as observed in the 2D simulations of Papers I 
and II: the generation of a hot shear layer around the core in those models in which the resonant modes already 
dominate during the linear regime (model B20), due to the strong dissipation of kinetic into internal energy in this area, 
when the resonant modes achieve the nonlinear regime. The right panels show the tracer, which represents the percentage of 
initial jet material in a cell. In the case of model B05, the whole jet is mixed with the ambient medium, as indicated by the 
maximum original jet mass fraction in a cell ($\sim$21 \%), model D10 shows some mixing down to the axis (with maximum mean 
jet mass fraction $\sim79\%$ on the jet axis), whereas in model B20 the central spine remains unmixed and collimated. 
Regarding the width of the mixing layer, we observe that model B05 presents the wider mixing region, whereas D10 has the 
thinnest one. However, if we consider that D10 was stopped much before the other models and that, from 
Fig.~\ref{fig:mom} the expected evolution is close to that of B05, we expect that the mixing layer of 
D10 after some time should be wider than that of B20.

\section{Discussion}\label{disc}

\subsection{Effects of the numerical resolution}
In \cite{pe+04a} it was shown that the resolution used in the simulations is crucial to reproducing the linear growth rates 
of the modes in the linear regime. In particular, the resolution may affect the evolution of the shortest (resonant) modes, 
which can be damped by numerical viscosity, and also have implications in the nonlinear regime with respect to, e.g., mixing. 
Thus, we performed several tests to check the influence of the resolution on our results. First, 2D 
numerical tests of models B05 and B20 were performed in which we used the same resolution as in the 3D simulations, 
and compared them with the results obtained for cylindrical jets in 2D (see Appendix B in Paper I). In these low-resolution 
simulations, we only excited the pinching modes. In the case of B05, the disruption of the 2D jet is faster, both in the 
low-resolution and the high-resolution cases, than in the 3D case, because long-wavelength modes have the largest amplitudes 
at the end of the linear regime. In 3D, the growth of short-wavelength modes (helical and/or elliptic) may be responsible for 
the slower mixing process. Between the 2D simulations, the low-resolution one shows a slower mixing and momentum transfer to 
the ambient, implying that the resolution affects this process, as expected. In model B20, we identify the effects of resonant 
modes also in the low-resolution 2D simulation, which result in a slow transfer of axial momentum from the jet to the environment 
during more than $2000 \,R_j/c$ code time units. However, the amplitude of these resonant modes at the end of the linear regime 
is less than in the high-resolution case. This translates into a decrease in the Lorentz factor down to the jet axis and more 
transfer of momentum from the jet to the ambient in the low-resolution case, around an equivalent time to what is reached by the 
high-resolution simulation. We continued the low-resolution simulation further and found that the oscillations in the amplitude 
of all the (excited) modes close to their limit of growth \citep[see][and Paper I]{pe+04b} end up in a long wavelength disrupting 
the flow. This occurs at time $t\simeq 2500\,R_j/c$, which, for a flow with radius, e.g., $R_j=100\,\rm{pc}$ propagating close to 
the speed of light, implies covering a distance greater than $200\,\rm{kpc}$ before disruption. We should expect that higher 
resolution would allow for an undamped growth of resonant modes, thus making these jets even more stable. 

In addition, we performed a 3D simulation of model B05 with twice the resolution used here. In this case, we would expect 
a faster growth of the resonant modes already detected in the 3D low-resolution simulation. This simulation was run in the linear 
and postlinear regime only during $\simeq 150\,R_j/c$ code time units due to the computational time limitations. The results 
confirm the previous statements: the (helical or elliptic) resonant modes grow faster and make the mixing of this jet 
slower (up to the moment in which we stopped the simulation) when higher resolution is used. 

In conclusion, the low resolution may quantitatively affect the results, damping the growth of the resonant modes and also 
reducing the mixing and momentum transfer rates, but the results presented here are qualitatively equivalent to those obtained 
in 2D with higher resolution.

\subsection{Implications for jet collimation}

The results are thus consistent with those of Papers I and II, where the authors showed that the growth of resonant 
modes in sheared jets can influence the evolution of relativistic flows, by generating small-scale mixing in the shear layer and 
avoiding the generation of large scale nonlinear structures that may disrupt the whole jet. This result could be important for our 
understanding of the properties of stability of relativistic outflows on the kiloparsec scales. Even though the Lorentz factors of 
both FRI and FRII jets appear to be similar on the parsec scales \citep{gi01,cg08}, the present paradigm of FRI jets 
\citep{bi84,la93,la96} states that, on kiloparsec scales, these jets are decelerated by entrainment of gas and their 
properties become different from those of FRII jets on these scales. Two main processes have been invoked to explain the 
entrainment in jets and studied in the literature: (i) entrainment through mixing in a turbulent shear layer between the jet 
and the medium \citep{dy86,dy93,bi94}, and (ii) entrainment from stellar mass losses \citep{ko94,bo96,lb02}. Recent numerical 
simulations deal with this last process (Perucho et al., in preparation). Concerning the process of entrainment through a 
turbulent shear layer, \cite{pm07} showed, via a 2D axisymmetric simulation, that a recollimation shock in a light jet, in 
reaction to a steep interstellar density and pressure gradients, may trigger nonlinear perturbations that lead to jet mixing 
and disruption with the external medium. Later, \cite{ro08} obtained a similar result in 3D for a homogeneous ambient medium, 
where the shock is induced by an overpressured cocoon. Also \cite{me08} studied the influence of a discontinuous jump in 
the external density gas on the jet disruption in hybrid FRI/FRII sources, where the jet and counterjet present different 
morphology, and this is adscribed to such environmental irregularities.

Regardless of the origin of the deceleration, the Lorentz factors in FRI jets after $1-2\,\rm{kpc}$ may already be small enough 
to favor the growth of KH modes, triggering the mixing process. In the context of the FRI/FRII dichotomy, the evolution of 
the KH instability can thus be seen as a consequence of this dichotomy and a cause of the kiloparsec-scale morphology of the 
two types of jets, where in one case the growth of the KH modes favor mixing and in the other may favor further collimation. 
In the case of FRII jets the strong recollimation shocks or strong entrainment discussed in the previous paragraph do not seem 
to happen. Nevertheless, they do show knotty structure due to oscillations around pressure equilibrium with the ambient 
medium or to travelling perturbations. In both cases and also if perturbations are induced by irregularities in the 
surrounding medium \citep[typically the backflow or cocoon, e.g.,][]{mi10}, these could couple to KH instabilities 
\citep[e.g.,][]{ag01}. Despite this, the FRII jets seem to be collimated up to distances of hundreds of kiloparsecs. 
To our knowledge, there are only two proposed mechanisms that could keep jets collimated: (i) the action of magnetic fields, 
a mechanism which does not work if the jets are not magnetically dominated, as is thought to be the case at kiloparsec 
scales \citep[see, e.g.,][]{sk05}; or (ii) by inertial confinement, which seems not to be possible within the low-density 
cavities with decreasing pressure \citep[see, e.g. simulations by][]{sc02,pm07}, where large-scale jets propagate. 
We propose here a third mechanism, purely hydrodynamical, that prevents the decollimation of relativistic jets, 
based on the growth of resonant modes.

According to our results (Papers I and II and this work), this mechanism is more effective in jets with larger flow Lorentz 
factors, but also depends on the initial width of the shear layer and on the temperature of the jet. Solutions to the linear 
perturbation problem in a sheared, relativistic jet show that, if the jet is separated from the ambient by a pure contact 
discontinuity (vortex sheet) or by a very thick shear layer ($> 1\,R_j$)\footnote{Taking the jet radius as the distance where 
the velocity is half that in the spine, typically $\simeq 0.5\,c$}, the resonances do not develop, thus leading to unstable 
configurations in which the jets can be disrupted by the growth of instabilities. The same solutions, along with numerical 
simulations, show that the resonances grow faster in colder jets, at a constant Lorentz factor. In addition, it was shown in 
previous works that lower density ratios between the jet and the ambient medium may also lead to the disruption of the jet 
by the growth of instabilities. 

We know at least one case in which a jet classified as an FRII \citep[0836+710,][]{hu92} for its luminosity, seems to be 
disrupted by the growth of instabilities on the kiloparsec scales. This could be assigned to the growth of instabilities 
\citep{pel07}. Perhaps in this case, the conditions in the jet (e.g., thickness of the shear layer or physical parameters 
in the jet) are such that resonant modes are not excited.

One observational feature of the growth of these modes would be the edge brightening of jets, which should mainly occur in 
FRII jets. Possible evidence of hot shear layers is given by Swain, Bridle \& Baum (1998), who show
that the jets in the FRII radio galaxy 3C353 have some transversal structure with higher radio-emission towards the boundaries 
of the jet. These hot shear layers could be generated by the growth of resonant modes as shown by our simulations. 
In addition, \cite{kad08} suggest the growth of resonant modes as the mechanism that provides the long-term stability of the 
jet in the FRII radio galaxy 3C111. A possible test for this hypothesis would be to check the transversal structure in this 
and other FRII jets seeking enhanced emission at the jet boundaries. From the point of view of particle acceleration, 
the hot and turbulent (down to numerical resolution) mixing layers produced by the resonant modes may be sites of efficient 
acceleration of non-thermal particles \citep[e.g.,][]{so02,am08}.

\section{Conclusions}\label{conc}
In this work, we presented the first solutions of the linear problem for 3D sheared relativistic flows. 
Following this result, we presented numerical simulations of the growth of KH instability modes from the linear to the nonlinear 
regimes, for three selected relativistic jet models. These numerical simulations were performed with the new high-resolution 
shock-capturing 3D RHD code \textit{Ratpenat}, which was parallelized using a hybrid scheme with OMP and MPI processes. 
The simulations of jets in the typical stability regions of the space of parameters (Paper~I) have confirmed previous results 
obtained in the 2D approximation. We proved that under certain conditions, the appearance of the shear-layer resonant modes 
of KH instability has important consequences for the stability of relativistic jets. In particular, the implications for 
the collimation of FRII jets were discussed. We also show, as a by-product of our simulations, that the development of 
different modes with a variety of wavelengths may induce rich transversal structure in jets, with shorter modes showing up in 
the central regions of the jet and longer ones dominating the structure in the mixing layer. In this respect, we confirm 
the results obtained in \cite{pe+06}.    

Future work along this line will consider the extension of our study in both the linear and nonlinear regime by means of 
simulations with higher resolution to a wider sample of jet parameters, including magnetic fields and a more realistic 
equation of state.

%\section*{Acknowledgements}
\begin{acknowledgements}
MP acknowledges support from a postdoctoral fellowship of the ``Generalitat Valenciana'' 
(``Beca Postdoctoral d'Excel$\cdot$l\`encia'') and a ``Juan de la Cierva'' contract of the Spanish 
``Ministerio de Ciencia y Tecnolog\'{\i}a''. MP and JMM acknowledge support by
the Spanish ``Ministerio de Educaci\'on y Ciencia'' and the European Fund for Regional Development 
through grants AYA2007-67627-C03-01 and AYA2007-67752-C03-02 and Consolider-Ingenio 2010, ref. 20811. 
MH acknowledges support by Polish MNiSW  grant 92/N-ASTROSIM/2008/0. MP benefitted from the HPC-Europa pogramme. 
The authors acknowledge the BSC. We thank E. Ros for criticism and comments to the original paper.
\end{acknowledgements}

%______________________________________________________________

\end{document}